\documentclass[useAMS,usenatbib]{mn2e}

\title[Tidal tails in galaxies]
{The symmetries and scaling of tidal tails in galaxies}
\author[C. Struck and B J. Smith] 
{Curtis Struck,\thanks{E-mail: curt@iastate.edu (CS);
smithbj@etsu.edu (BJS)}$^1$
Beverly J. Smith$^{2}$ \\
$^1$ Dept. of Physics and Astronomy, Iowa State Univ., Ames, IA, 50011 USA\\
$^2$ Dept. of Physics and Astronomy, East Tennessee State Univ., Johnson City, TN, 37614 USA}
\usepackage{amssymb}
\usepackage{amsmath}
\usepackage{graphicx}
\usepackage{epsfig}
\usepackage{multirow} 
\def\aap{{ A\&A}}

\def\aj{{AJ}}

\def\apj{{ApJ}}

\def\mnras{{MNRAS}}

\def\nat{Nature}
\def\ARAA{{ARA\&A}}

\begin{document}
\date{\today}

\pagerange{\pageref{firstpage}--\pageref{lastpage}} \pubyear{0000}

\maketitle

\label{firstpage}
\begin{abstract}
We present analytic models for the formation and evolution of tidal tails and related structures following single or multiple impulsive disturbances in galaxy collisions. Since the epicyclic approximation is not valid for large radial excursions, we use orbital equations of the form we call p-ellipses (a class of precessing ellipses). These have been shown to provide accurate representations of orbits in logarithmic and power-law halo potentials. 

In the simplest case of an impulsive collision yielding a purely tidal disturbance the resulting tidal tails have simple structure. Scalings for their maximum lengths and other characteristics as nonlinear functions of the tidal amplitude and the exponent of the power-law potentials are described. The analytic model shows that azimuthal caustics (orbit crossing zones of high density also seen in numerical models) are produced generically in these tails at a fixed azimuth relative to the point of closest approach. Long tails, with high order caustics at their base, and ocular waveforms are also produced at larger amplitudes. 

The analysis is then extended to nonlinear disturbances and multiple encounters, which break the symmetries of purely tidal perturbations. The p-ellipse orbital solutions are similar to those in the linear tidal case. However, as the strength of the nonlinear terms is varied the structure of the resulting forms varies from symmetric tails to one-armed plumes. Cases with two or more impulse disturbances are also considered as the simplest analytic models distinguishing between prograde and retrograde encounters. The model shows explicitly how tail growth differs in the two cases. In the prograde case a specific mechanism for the formation of tidal dwarf galaxies at the end of tails is suggested as a consequence of resonance effects in multiple or prolonged encounters. Qualitative comparisons to Arp Atlas systems suggest that the limiting analytic cases are realized in real systems. For example, we identify a few Arp systems which have multiple tidal strands meeting near the base of long tails. These may be swallowtail caustics, where dissipative gas streams are converging and triggering star formation. UV and optical images reveal luminous knots of young stars at these 'hinge clump' locations. 
 
\end{abstract}


\section{Introduction}
Spiral waves, and their longer extensions which we now recognize as tidal tails, have fascinated since the studies of \citet{ho40, ho41}, \citet{zw59} and others. Holmberg and Zwicky speculated that tails, and related structures, were produced by tidal forces in galaxy collisions. This was well confirmed by the numerical case studies described by \citet[with references to other modeling efforts]{to72}. These results have in turn been amply confirmed by subsequent generations of increasingly comprehensive numerical simulations, including detailed studies of specific systems (see e.g., the recent papers of \citealt{st05}, \citealt{sm08}, \citealt{ch10}, \citealt{ba11}, and references therein), and studies of detailed features that can develop within the tail, like tidal dwarf galaxies (e.g., \citealt{ba92b}, \citealt{el93}, \citealt{du04}, \citealt{bo06}, \citealt{bo08}). The now extensive literature of numerical models provides a broad understanding of the basic processes of tail formation and evolution. However, as usual with simulations, it is hard to grasp all the systematics of a complex phenomenon like tidal tails even with the results of many of specific calculations available. Analytic models or scalings are often very helpful for understanding such trends.

A number of papers in the literature have used analytic or semi-analytic models to study waves produced in galaxy collisions. One of the first of these was Toomre's (see \citealt{to78}, also \citealt{st90b}) illustration of the radial orbital motions that make the circular waves of a colliding ring galaxy. Later came a number of analytical attempts to map out less symmetric evolving wave structures through disturbed discs using the Impulse Approximation for the perturbation, and the epicyclic approximation for initially circular orbits following the perturbation (\citealt{st90a}, \citealt{do91}, \citealt{ge94}). In these studies the orbits are approximated analytically, but the waves were studied by numerical calculations of the post-collision orbital motions of sample particles in a model disc. The degree to which such models can match more sophisticated, self-consistent simulations is generally impressive at early stages. However, these initial collisions are often very impulsive, and the full simulations are required to model detailed dynamics of the final mergers of galaxies (e.g., reviews of \citealt{ba92a}, \citealt{st99}). This limitation of the semi-analytic models is not surprising. More vexing was the fact that the evolving wave structures are complex, and seem to depend sensitively on values of collision parameters. Thus, it was not immediately obvious that the semi-analytic models could provide any more predictive power than case-by-case studies with simple simulations, e.g., restricted three-body calculations. 

However, there is perhaps more hope for analytic models of the simplest and most symmetric galaxy collisions, as in the case of symmetric ring galaxies. In a series of papers, one of us has derived a complete analytic model of stellar ring waves, subject to the Impulse and epicyclic approximations, and the further approximation of rigid halo potentials (\citealt{ap96}, \citealt{st10}). The key to this work was the realization that the orbit-crossing zones that define the waves can be described with the mathematical machinery of caustic or catastrophe theory \citep{ar86}. 

More recently, we have been able to extend this analytic theory to cases of spiral waves produced in impulsive flyby encounters \citep{st11}. This theory applies in cases where the galaxy discs have a relatively high value of the Toomre Q parameter, so that the induced spirals are not strongly self-gravitating, and the epicyclic approximation can be used to describe the post-encounter orbits. The obvious difference from the ring galaxies is that the magnitude and direction of the velocity impulse, and thus, the parameters describing the epicycles, vary with azimuth. The radially and azimuthally dependent velocity impulses can be simply approximated by classical tidal formulae, or more accurately in specific cases by formulae like those derived, for example, by \citet{do10}. 

The first difficulty that confronts anyone wanting to extend such analytical models to tidal tails is their length. Tidal tails can stretch out to distances beyond twice the original orbital radius of the stars or gas clouds within them. Using a single epicycle to approximate disturbed orbits works best when the disturbance is relatively small, and the radius of the epicycle is a fraction of the original, or the guiding centre, radius. Thus, the epicycle approximation applied to tail particles is an absurdity. In such encounters, stars in two disc quadrants have their angular momentum increased, and fly out in bridges or tails. Those in the other two quadrants lose comparable amounts of angular momentum, and in simulations these stars generically pile up in the central disc. The epicyclic approximation cannot represent this phenomenon well either.

Another problem in analytically modeling tails is that the conventional wisdom has it that long tails cannot be produced in impulsive encounters; more torque is needed to pull tail particles far from the disc centre. Specifically, the Impulse Approximation does not distinguish between prograde and retrograde encounters, and \citet{do10} argue that it is nearly resonant encounters in the former case that produce long tails. These authors compute the perturbation to circular stellar orbits due to the finite-time pull of a companion on a fixed orbit. They suggest that one can assume that these (for some stars semi-resonant) perturbations are produced instantly in a kind of semi-impulse approximation, which would distinguish between prograde and retrograde encounters. This work has similarities to the Linear Response Theory approach of \citet{ne99}. Both provide means of improving the classical Impulse Approximation. In the present work we will use the classical approximation, with some extensions. We will also probe the limits of the generalization that tails cannot be produced in impulsive encounters. 

But how should we produce such analytic models without the simple and versatile epicyclic orbit approximation? We are forced to look at other types of orbits, and a variety of other forms have been suggested, even in recent years. All of them ultimately have the approximate form of precessing ellipses. However, \citet[henceforth S06]{st06}, \citet[henceforth LBJ]{ly08} and \citet{ly10a} rediscovered a particular class, whose members are described by simple formulae, and which provide good fits to orbits in power-law and logarithmic potentials, ranging from nearly circular to nearly radial. The ÔfitsÕ are very good for near circular orbits. They allow errors at second-order in the eccentricity, but since they accurately approximate the precession period, they do not ÔdriftÕ in phase relative to the true orbit. (Better, but more complex approximations were also explored in S06.) The symmetry properties of these curves, which we call p-ellipses (for precessing ellipses in power-law potentials), have been explored by LBJ. The Impulse Approximation, and orbits described by these curves provide the basis for constructing analytic models of tails, as we will shortly describe (also see \citealt{ly10b}). 

Before beginning that discussion we note that we will restrict attention to free tails without connection to a companion. Bridges experience prolonged forcing from the companion, so we cannot expect the Impulse approximation to apply. The models discussed below do not include the effects of self-gravity in disc waves, and are two-dimensional. That is, the orbital plane of the companion is assumed to be the same as the disc plane of the target, and there is no vertical perturbation. It would be straightforward to generalize the models to include moderate vertical effects, but we do not do so in this paper.

\section{Analytic model equations and their properties}

\subsection{Post-impulse p-ellipse orbits}
To begin the analytic study we assume, for simplicity, that the disc orbits are circular before the encounter. The polar equation for p-ellipse orbits can be written, 

\begin{equation}
\frac{p}{r} =  \left[ 1 +
e \cos \left( m{\theta} \right) 
\right]^{\frac{1}{2} + \delta},
\end{equation}

\noindent
where the orbit of an individual star is specified by the semi-latus rectum $p$, which sets the scale of the orbit, the eccentricity $e$, and an initial value of the phase $\theta_o$. The parameter $\delta$ is determined by the form of the gravitational potential, e.g., $\delta = 0$ for the logarithmic potential, and $\frac{1}{2}$ for the Kepler potential. The factor $m$ is a simple function of $\delta$, see S06 for details. (In S06 this factor was written as $(1-b)$, but here we use the simpler notation of LBJ.)

Assuming that the form of the potential is fixed, after the impulse the problem of determining the p-ellipse orbit of a star is equivalent to finding the values of $p$, $e$, and $\theta_o$ (which determines the initial position of the star on the p-ellipse). A star's position in the disc is specified by radius $r$, and azimuth, $\phi$. We denote the coordinates of a star at the time of impact as $r = q$, and $\phi = \phi_o$. In the Impulse Approximation these are unchanged immediately after the impact; only radial and azimuthal velocities are changed. We choose the coordinate system such that the impact occurs at azimuth $\phi = 0$. Equation (1) shows that when the p-ellipse azimuth $\theta = 0$, $r$ takes its minimum value, which does not generally occur at $\phi = 0$. Thus, the two azimuthal coordinates are offset, and in general $\theta_o \ne \phi_o$ for any star. Note that we use the symbol $\theta$ for the p-ellipse azimuth in this work, while $\phi$ was used in S06.

Let us return to the issue of finding relations that can be solved for the values of $p$, $e$, and $\theta_o$ for a stellar orbit. The first is given by equation (1), written in terms of initial values,

\begin{equation}
\frac{p}{q} =  \left[ 1 +
e \cos \left( m{\theta_o} \right) 
\right]^{\frac{1}{2} + \delta},
\end{equation}

\noindent To obtain a second relation we differentiate equation (1) to get a radial velocity equation. This can be written as, 

\begin{equation}
\frac{p{\dot{r}}}{r^2} =  \frac{\left(\frac{1}{2} + \delta\right) me   
{\sin \left( m{\theta} \right)} \dot{\theta}}
{\left[ 1 + e \cos \left( m{\theta} \right) 
\right]^{\frac{1}{2} - \delta}},
\end{equation}

\noindent
where ${\dot{r}}$ and $\dot{\theta}$ are the time derivatives of the coordinates, i.e., the radial and angular velocities. The initial values of these velocities are determined by the Impulse Approximation. Equation (1) can be used to eliminate $p/r$, and equation (3) can be solved for the ratio of velocities, and evaluated at the time of impact as a function of $q$ and $\theta_o$.  I.e., 

\begin{equation}
\frac{v_r}{v_{\theta}} =  \frac{\left(\frac{1}{2} + \delta\right) me 
\sin \left( m{\theta} \right)}
{\left[ 1 + e \cos \left( m{\theta} \right) \right]},
\end{equation}

\noindent
where $v_r = {\dot{r}}$, and $v_{\theta} = q\dot{\theta}$. Note that due to fortuitous cancellations there are no exponents with factors of $\delta$ in equation (4).

Equation (4) is the second relation between the p-ellipse parameters.  A third relation can be obtained from the conservation of angular momentum, and these equations can then be solved for the three orbital parameters. It turns out, however, the symmetries of simple tidal perturbations allow us to avoid this latter calculation.

\subsection{Simple tidal perturbations}

As noted above, the encounter perturbation can be approximated at various levels of accuracy. \citet{do10} give some of the most sophisticated analytic formulae published to date, and they are quite complex. \citet{ge94} computed the nonlinear impulse from a Plummer potential perturber on a parabolic orbit, which is less complex, though still not simple. When combined with the equations for the p-ellipse orbits, complex expressions for the perturbation velocities can render further calculations very difficult. To begin to explore what can be learned from analytic models, we begin with the lowest order impulse formulae derived from equation (7) of Gerber and Lamb. We call this the ÒPure TidalÓ case. These impulses can be written, 

\begin{equation}
{\Delta}{v_r} = A \left( \frac{q}{r_{min}} \right) 
v_{{\phi}o} \cos \left( 2{\phi}_o \right),
\end{equation}

\begin{equation}
{\Delta}{v_{\phi}} = -A \left( \frac{q}{r_{min}} \right) 
v_{{\phi}o} \sin \left( 2{\phi}_o \right),
\end{equation}

\noindent with,

\begin{equation}
A = \frac{2GM_c}{r_{min}Vv_{{\phi}o}}.
\end{equation}

\noindent In these expressions $r_{min}$ is the distance of closest approach between the galaxy centres, $V$ is the relative velocity at closest approach, $M_c$ is the total mass of the companion. Equations (5) and (6) are the approximation of Gerber and Lamb's equation to first order in the quantity $q/r_{min}$, which are nonlinear in this ratio (see Section 5). Note that these expressions differ from textbook forms for the tidal accelerations in the Earth-Moon system and other published forms (e.g., the factor of 2 in the argument of sine and cosine). These differences trace to different assumptions about the acceleration of the centre of the disc galaxy. 

\subsection{Orbital parameters}

Equations (5) - (7) can be combined to obtain an expression for the post-impulse velocity ratio, analogous to equation (4),

\begin{equation}
\frac{v_r}{v_{\phi}} =  
\frac{{\Delta}{v_r}}{{v_{{\phi}o}} + {\Delta}{v_{\phi}}} =
\frac{A' \cos \left( 2{{\phi}_o} \right)}
{ 1 - A' \sin \left( 2{{\phi}_o} \right)}.
\end{equation}

\noindent In this equation we use the radius-dependent amplitude,

\begin{equation}
A' = A \left( \frac{q}{r_{min}} \right).
\end{equation}

\noindent The form of equations (4) and (8) is very similar. As discussed in Sec. 2.1, we expect to fit a post-collision orbit with a p-ellipse. We can do this by setting equations (4) and (8) equal to each other,

\begin{equation}
\frac{\left(\frac{1}{2} + \delta\right) me 
\sin \left( m{\theta}_o \right)}
{\left[ 1 + e \cos \left( m{\theta}_o \right) \right]} = 
\frac{A' \cos \left( 2{{\phi}_o} \right)}
{ 1 - A' \sin \left( 2{{\phi}_o} \right)},
\end{equation}

\noindent and requiring that the numerators and denominators (radial and azimuthal velocities) on each side are equal. 

The solution is obtained by taking,

\begin{equation}
m{\theta}_o = 2{\phi}_o + \frac{\pi}{2} + \psi,
\end{equation}

\noindent where the ${\pi}/2$ term turns sine into cosine and vice versa, and the angle $\psi$ gives the remaining phase difference. With this relation between the angles, we equate the numerators and denominators of equation (10), and solve for $A'$ and $\psi$. We get,

\begin{equation}
A' = \frac{\left(\frac{1}{2} + \delta\right) me}
{\sqrt{\left(\frac{1}{2} + \delta\right)^2  m^2
\sin^2 \left( 2{{\phi}_o} \right) +
\cos^2 \left( 2{{\phi}_o} \right) }},
\end{equation}

\noindent and,

\begin{multline}
\tan \left( 2{\phi}_o + \psi \right) = 
{\left(\frac{1}{2} + \delta\right)} m \tan \left( 2{\phi}_o \right) = \\
{\left(\frac{1}{2} + \delta\right)} \sqrt{2 \left( 1 - \delta \right)} 
\tan \left( 2{\phi}_o \right).
\end{multline}

\noindent where the last equality in equation (13) makes use of the first order approximation from S06 for $m(\delta)$. Equation (12) gives $A' \simeq e$ for most values of $\phi_o$. Thus, the eccentricity of the p-ellipse depends weakly on the initial azimuth. The equations also show that perturbation amplitudes of order unity are required to get high eccentricities, and nearly radial orbits. 

The solution to equation (10) given by equations (11)-(13) is very simple. Note that the use of equation (11) in equation (1) also yields an expression for the semi-latus rectum of the p-ellipse,

\begin{multline}
\left( {\frac{p}{q}} \right)^{\frac{1}{\frac{1}{2} + \delta}} = 
1 + e\cos \left(2{\phi_o} +\frac{\pi}{2} + 
\psi \right) = 1 -  \\
\left( 
\frac{\sqrt{\left(\frac{1}{2} + \delta\right)^2  m^2 
\sin^2 \left( 2{{\phi}_o} \right) + \cos^2 \left( 2{{\phi}_o} \right) }}
{\left(\frac{1}{2} + \delta\right) }
\right)
A' \sin \left(2{\phi_o} + \psi \right).
\end{multline}

\noindent This parameter depends on both the star's initial radius and azimuth.

\subsection{Accuracy of the approximation}

The obvious question about the simple approximate orbits of the previous subsection is how accurate are they, even within the assumed fixed potential of the parent galaxy? The immediate answer is that p-ellipse orbits in a power-law potential have been shown to be quite accurate (S06, LBJ, \citealt{ly10a}). Though initiated by an impulsive disturbance, the orbits described above will inherit those general characteristics. For example, one of the most remarkable properties of p-ellipse orbit approximations is that in many cases the fit to the true orbit does not deteriorate with time. That property also continues to hold. On the other hand, while the above p-ellipse approximation is quite accurate for low to moderate eccentricity orbits, it is significantly less accurate at higher eccentricities. Tidal tails are made of quite eccentric orbits, so we will consider this issue more carefully below. 

We begin that discussion with Figure~1, which compares 11 numerically computed orbits, each at a different initial azimuth relative to the impulse impact azimuth, and approximate orbits given by the equations of the previous subsection in the logarithmic potential ($\delta = 0$). The initial radius for all the orbits shown is the outer disc radius, which is scaled to 1.0. In the graph, successive orbit pairs are offset by one unit for clarity. The impulse amplitude in the case shown in Figure 1 has a relatively high value of $A'= 0.5$. Nonetheless, the numerical orbits (solid curves) are quite well fit by the analytic approximation (dashed curves). The maximum deviations in radius between the two curves are about 50\% (of the larger radius), and generally much less.

\begin{figure}
\centerline{
\includegraphics[scale=0.4]{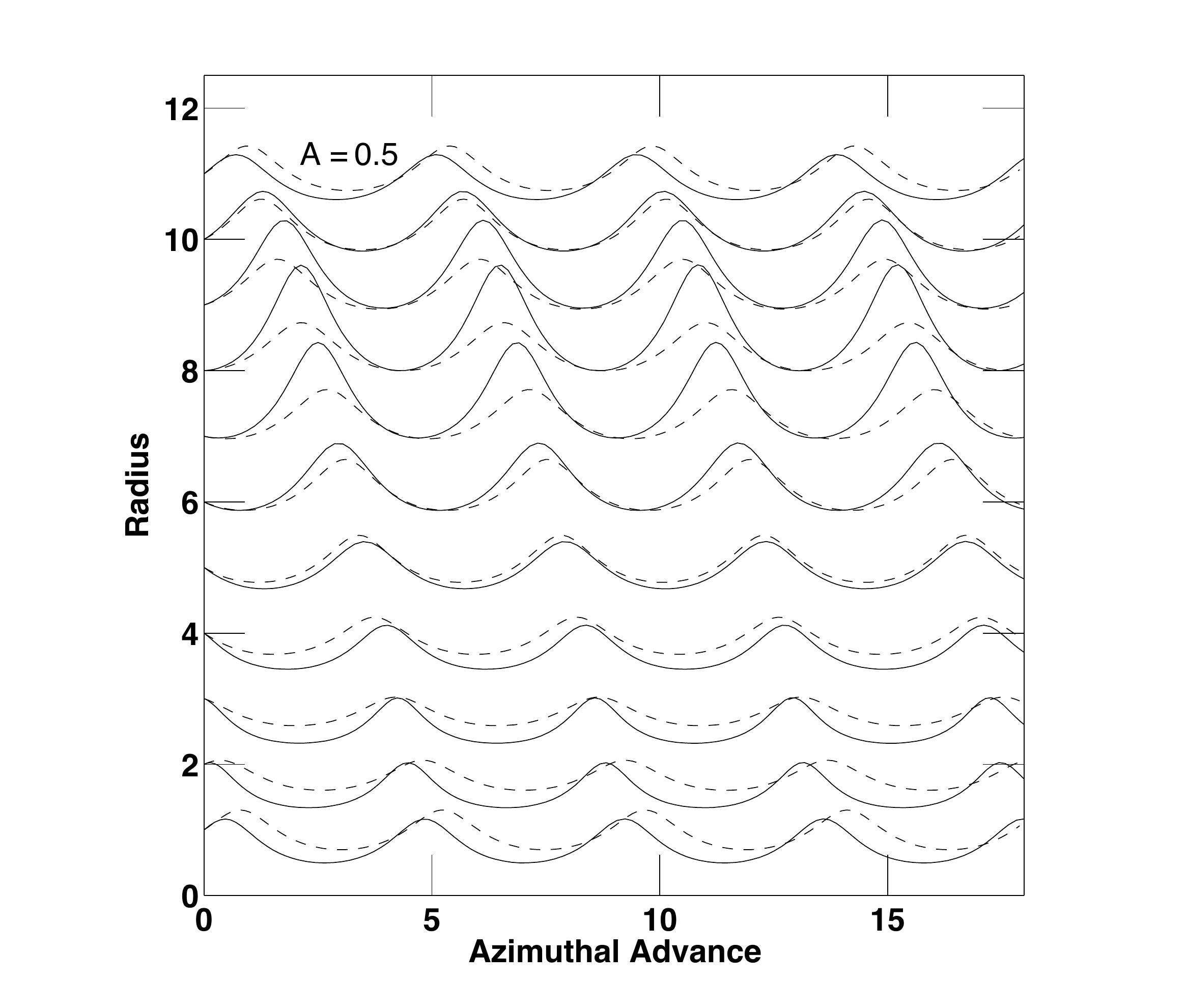}}
\caption{Radius versus azimuthal increase from the initial value (as a proxy for time) is shown for sample numerically integrated (solid curves) and analytically approximated (dashed curves) stellar trajectories. The initial radius in all cases is $q = 1.0$, but successive pairs of curves are offset by one unit in radius for graphical clarity. Successive pairs of curves are distinguished by different initial azimuths relative to the line connecting the disc centre to the point of closest approach (i.e., the impulse direction). From bottom to top the initial azimuths are: 0.3, 0.6, 0.9, 1.2, 1.5, 1.8, 2.1, 2.4, 2.7, 3.0, 3.3 radians. The form of the potential is logarithmic ($\delta = 0$) and the amplitude of the tidal impulse is $A'= 0.50$ (at $q = 1.0$) in all cases. In the analytic approximation the eccentricities were computed using equation (12).}
\end{figure}

Within this figure it can be seen that the fit is better for some trajectories than for others. The fit correlates directly with eccentricity, which correlates with amplitude according to equation (12), but also with initial azimuth via the tidal perturbation. Higher values of the eccentricity in the post-impulse orbit yield worse fits. The case shown in Figure~1 is in fact a borderline case. The fit is generally much better for low amplitude cases, and gets worse rapidly as the amplitude increases above 0.5. This can be seen in Figure~2, which shows a higher amplitude case. This is easy to understand. The analytic solution solves the equation of motion to order $e$. Beyond $A', e \simeq 0.5$, the effect of the nonlinear terms grows rapidly. 

\begin{figure}
\centerline{
\includegraphics[scale=0.4]{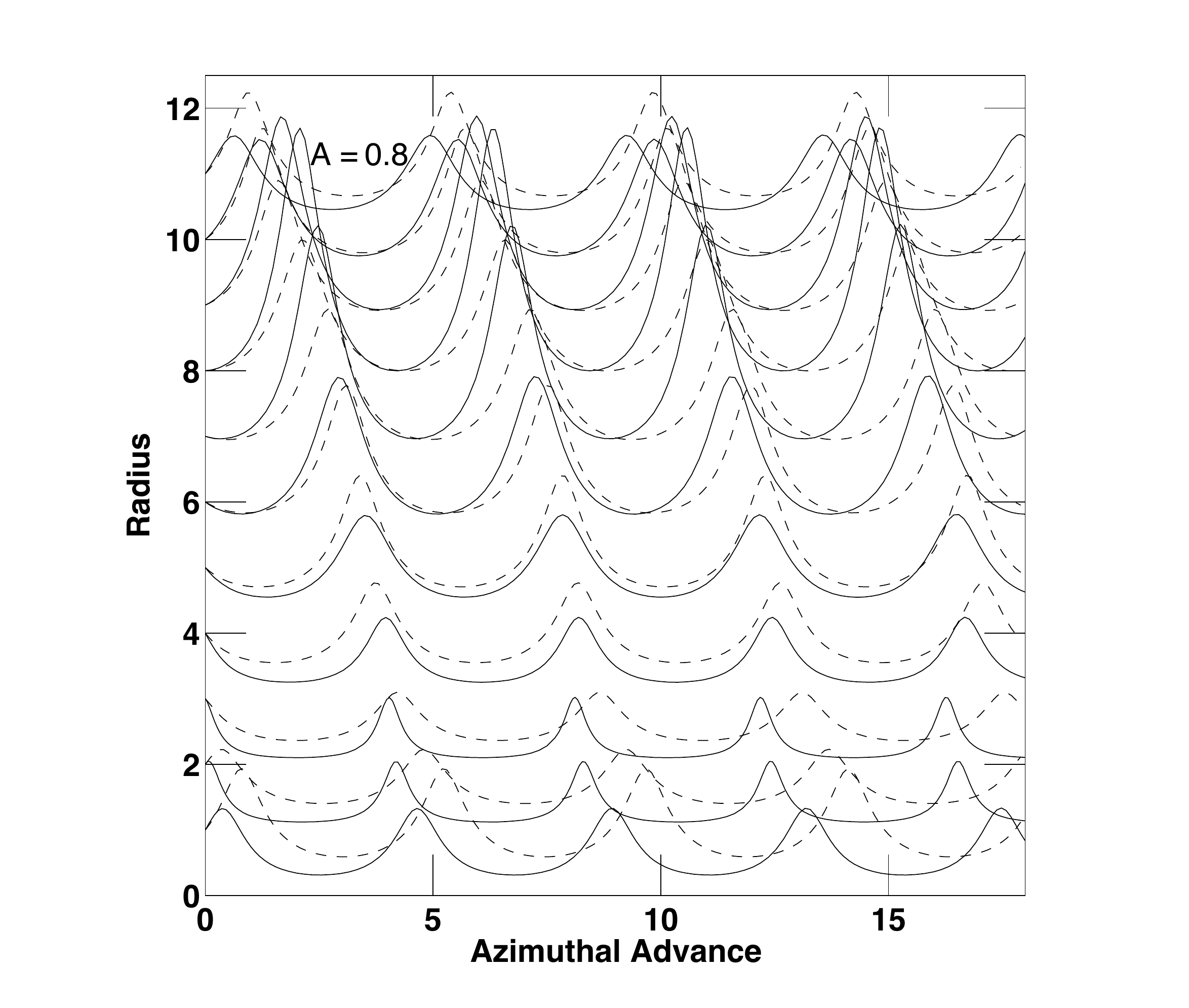}}
\caption{Same as Figure 1, but with an impulsive tidal amplitude of $A'= 0.80$ (at $q = 1.0$).}
\end{figure}

The fact that a simple p-ellipse approximation works up to amplitudes as high as 0.5 is impressive. At somewhat higher amplitudes, the eccentricities of specific orbits can exceed values of 0.9, and the difference between numerical and analytic curves can exceed factors of a few. It is true that only a small fraction of all orbits are affected, and they could be neglected, but this fraction grows with amplitude. Thus, the approximation becomes increasingly less useful.

Fortunately, a little further analysis allows us to largely overcome this limitation. The inaccurate analytic eccentricities occur when the cosine term in equation (12) is large. An ad hoc, but modest change of that equation, to the form, 

\begin{equation}
A' = \frac{\left(\frac{1}{2} + \delta\right) me}
{\sqrt{\left(\frac{1}{2} + \delta\right)^2  m^2
\sin^2 \left( 2{{\phi}_o} \right) +
\left( \frac{1}{1+A'} \right)
\cos^2 \left( 2{{\phi}_o} \right) }},
\end{equation}

\noindent proves very helpful. The added factor of $1/(1+A')$ largely cancels the effects of the nonlinear terms, and the modified analytic orbits fit the numerical ones quite well. This is true throughout the range $0.5 \le A' \le 0.9$, while the effect of this extra factor is negligible at small amplitudes. With this modification we have a qualitatively good orbit approximation over most of the interesting range of amplitude. 

The examples shown in Figures~1 and 2 are for the $\delta = 0$ case. The improved approximation of equation (15) does not work as well for larger (or smaller, negative) values of $\delta$. In the Keplerian case ($\delta = 0.5$), it starts to break down for amplitudes greater than about 1/3. Specifically, the maximum radial excursions of some of the numerical orbits grow very large as the amplitude increases. This is not a surprise. An azimuthal pertubation amplitude of $2^{1/2} - 1$ boosts the stellar velocity to the classical escape velocity. There is no such problem in the logarithmic potential, but there will be in all the potentials with values of $\delta$ greater than zero.

Equation (15) is designed to tame the excessive excursions in the analytic orbits at high amplitude in the $\delta = 0$ case, and a similar modification of the sine term in that equation can boost the analytic solution at the correct phases to match the larger numerical excursions in the Keplerian and intermediate cases. For example, the form,

\begin{multline}
A' = \\
\frac{\left(\frac{1}{2} + \delta\right) me}
{\sqrt{ b_2
\left(\frac{1}{2} + \delta\right)^2  m^2
\sin^2 \left( 2{{\phi}_o} \right) + b_1
\cos^2 \left( 2{{\phi}_o} \right)}}\\
\mbox{with, } b_1 = \left( \frac{1}{1+A'} \right), 
b_2 = \left( 1+A' \right) ^{0.5 + 24{\delta}^2},
\end{multline}

\noindent yields fits better than or comparable to those of Figures~1 and 2 for all values of $\delta$ between 0 and 0.5, and for all, except the highest values of amplitude. This is shown in Figure~3, which is the same as Figure~2 except that the eccentricities are derived from equation (16) rather than equation (12). The effect is even more pronounced for cases when $\delta >  0$. The new coefficient of the sine term was derived by experimentation. It has not been proven to be an optimal fit, but it is a good one. The 0.5 term in the exponent of $b_2$ offsets much of the difference between analytic and numerical curves at $\delta = 0$ (see Figs. 1 and 2). A nonlinear dependence on $\delta$ is needed in the exponent to fit the range of potentials, and the large coefficient is needed to offset the low numerical value of $\delta^2$. 

\begin{figure}
\centerline{
\includegraphics[scale=0.4]{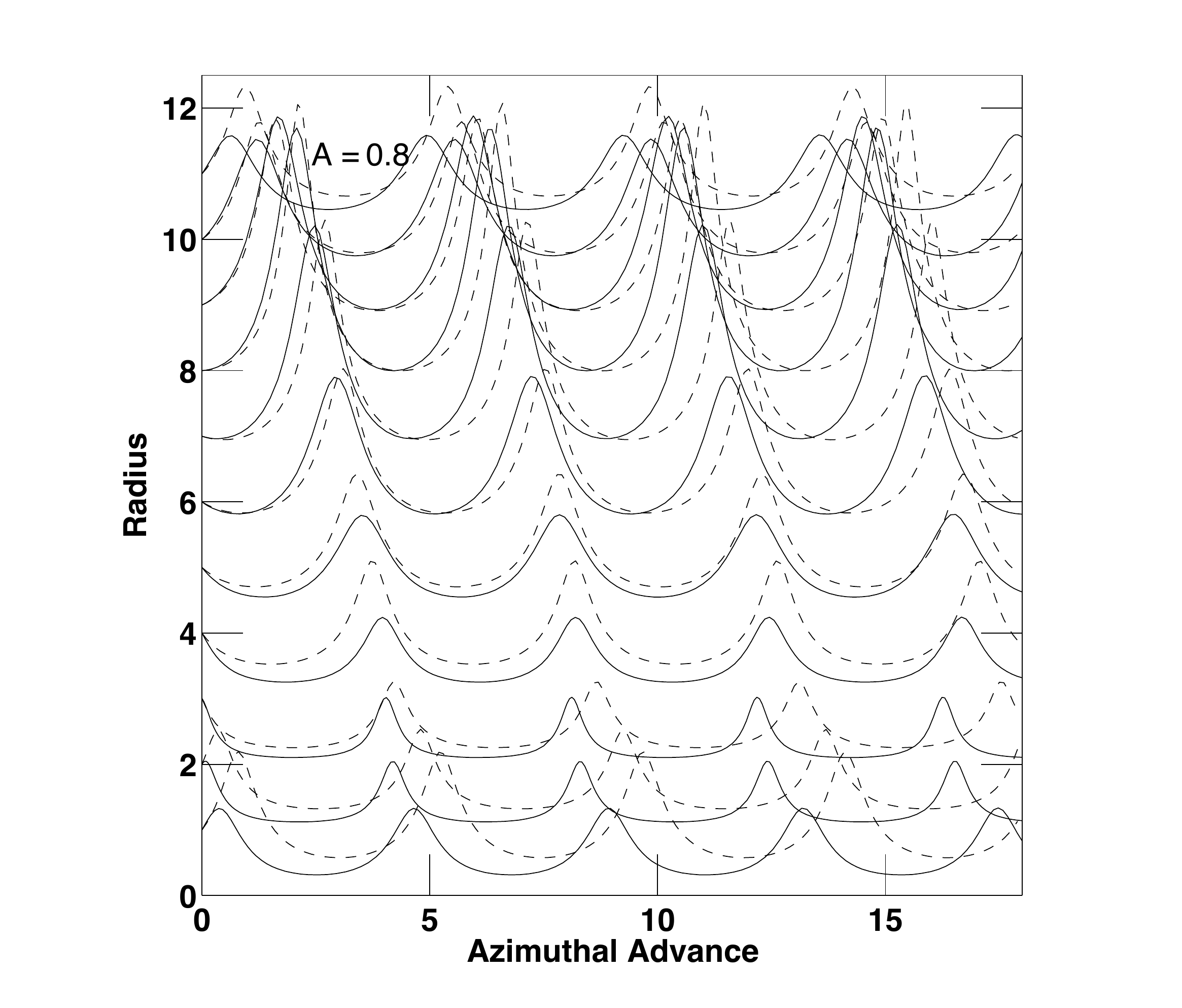}}
\caption{Same as Figure 2, except that in the analytic approximation the eccentricities were computed using equation (16). Comparison with Figure 2 shows the improvement, see text.}
\end{figure}

It is worth recalling that the refinements of the simple orbit approximation of the previous subsection embodied in equations (15) and (16) only change the value of the eccentricity. They still yield a p-ellipse fit to a given orbit, but one that is better than the fit achieved by the velocity matching of equation (10) at certain azimuths. That velocity matching gives a p-ellipse solution that is guaranteed to fit the orbit to first-order in $e$. In sum, we have shown that there is also a p-ellipse, with a different eccentricity, that fits some perturbed orbits better. In the present context this may appear fortuitous, but in fact it is a corollary of the result that any orbit in these power-law potentials can be well fit by a p-ellipse (S06, LBJ). We also note that D. Lynden-Bell has explored extended versions of the p-ellipse functions that provide still better approximations at high eccentricity (private communication).  

\subsection{Special case: the backbone of the tail}

Tails are produced by the strongest azimuthal impulses, and equation (6) implies that these will occur at azimuths of $\phi_o = -{\pi}/4$ and $3{\pi}/4$, relative to the line connecting the disc centre to the point of closest approach. The second of these azimuths is the one associated with the counter-tail. The radial velocity impulse is zero at these azimuths. 

Of special interest is the structure of the p-ellipse orbits of stars initially located along this azimuth, which can be regarded as the backbone of the tail. Since the radial impulse is zero at this azimuth, equation (4) yields $\sin(m{\theta_o}) = 0$, $m{\theta_o} = 0$, or $2\pi$, and $\psi = 0$ according to equation (11). Substituting this value into equation (14) we get, 

\begin{equation}
\frac{p}{q} = 
\left(1+A'\right)^{\left(\frac{1}{2} + \delta\right)}.
\end{equation}

\noindent Thus, in going from values of $\delta = 0$ (logarithmic) to $0.5$ (Keplerian), the orbit scale $p$ goes from a weak square root to a linear dependence on $A'$. The difference is even greater if we use equation (16), with its extra terms, for the eccentricity in equation (14).

Using equations (1), (14) and (16), we obtain the following expression for the maximum radial excursion (apoapse) of a backbone star, 

\begin{multline}
\frac{r_{ap}}{q} = 
\left(\frac{1+A'}{1-e} \right)
^{\left(\frac{1}{2} + \delta\right)} = \\
\left(\frac{1+A'}{1-A'\left(1+A'\right)
^{0.25+12{\delta}^2}} \right)
^{\left(\frac{1}{2} + \delta\right)}.
\end{multline}

\noindent This equation has several interesting properties. The first is that the denominator goes to zero at a critical value of the amplitude $A' \le 1$, so the expression allows for very long tails. If the colliding galaxies are bound, it can be shown that the amplitude is equal to a factor of order unity times the ratio $qM_c/(r_{min}M(q))$, where $M_c$ is the mass of the companion, and $M(q)$ the mass of the parent galaxy within the star's orbit. If the two masses are comparable and the distance of closest approach not much greater than $q$, then the critical amplitude can be exceeded. E.g., a close encounter with a more massive (and compact) companion will produce strong perturbations, and according to these results could send backbone stars out to large distances. When $\delta > 0$ stars can escape the potential as noted above. (These conditions are apparently satisfied in Arp 105 and 174, for example, not to mention the globular cluster Palomar 10.) This situation would probably entail the near destruction of the smaller galaxy, but it does suggest that in some cases, impulsive encounters can produce long tails. 

In the case of weaker disturbances, equation (18) suggests that it would be hard to launch a backbone star out to more than a few times its initial radius. For example, with $\delta = 0$ (logarithmic potential), and $A' = 0.3$, then $r_{ap}/q \approx 1.4$. 

The dependence of equation (18) on the form of the potential is also interesting. If we take $\delta = 0.5$ and $A' = 0.3$, then $r_{ap}/q \approx 4.4$. Other factors being equal that implies much greater stretching of tails in Keplerian potentials than in flat rotation curve potentials (see Sec. 3). As the amplitude is increased to $A' = 0.4$, the tail length ratio increases greatly. \citet{du04} noted this effect in their large grid of tail-making galaxy collisions. They argued that magnitude of the stretching was a key factor in determining whether a tidal dwarf galaxy could form at the end of a tail. If the tail material is stretched too much it is much more difficult for local self-gravity to assemble a dwarf. The earlier simulations of \citet{mi98} also showed that dwarfs are easier to manufacture in more extensive halos. The equations above suggest scalings for the tail stretching, and the basis of a method for estimating other tail properties. 

Another aspect of the tidal stretching results from the radius dependence of the amplitude $A'$. In equation (17) this nonlinear dependence results in greater stretching of stars with larger initial radii. This global stretching means that kinematic pile-ups and radial caustics are not possible in impulsive tails, but must be the result of prolonged disturbances or local self-gravity.

\subsection{Fleshing out the backbone}

In this section we briefly demonstrate that post-collision orbits in nearby sectors are very similar to those on the backbone. Qualitatively we expect this, given the smoothness and continuity of the simple tidal perturbation, but it can be demonstrated more quantitatively. We begin with equation (12) for the eccentricity. (The results here will also hold for the more complex form of equation (16).) For sectors near the backbone, we assume that $\phi_o \approx 3{\pi}/4 + \epsilon$, where $\epsilon << 1$. Substituting this into equation (12), solving for $e$, and expanding in powers of $\epsilon$, we obtain,

\begin{equation}
e = A'
\left[1 - 2{\epsilon}^2 \left(1 - 
\frac{1}{\left(\frac{1}{2} - \delta\right)^2  m^2}
\right) \right].
\end{equation}

\noindent Thus, the eccentricity has only a relatively weak, second-order dependence on $\epsilon$.

A similar expansion of equation (15) shows that the semi-latus rectum also depends weakly on the azimuthal deviation $\epsilon$,

\begin{multline}
\frac{p}{q} = \left(1 - A'
\sin\left( \frac{3\pi}{2} + 2\epsilon \right)
\right) ^{\frac{1}{2}+\delta}\\
\approx \left( 1 + A' \right)^{\frac{1}{2}+\delta}
\left( 1 - \frac{2{\epsilon}^2}{1+A'} \right).
\end{multline}

\noindent The weak dependence on $\epsilon$ in both p-ellipse orbit parameters confirms our statement that adjacent orbits are like those in the backbone, and that the backbone represents the whole tail.

\subsection{A sharp wave in the tail}

Section 2.5 ended with the comment that tidal stretching rendered kinematic crowding and caustic wavefronts unlikely in the radial direction. This is not necessarily the case in the azimuthal direction. Technically, the caustic condition requires that the Jacobean matrix (which measures compression and shear) of the coordinates with respect to their initial values go to zero. We assume that the derivative of the radius with initial azimuth (${\partial}r/{\partial}\phi_o$) is negligible, i.e., nearly uniform stretching in azimuths near $\phi_o= 3{\pi}/4$, in accord with the results of the previous section. Then the condition for caustic compression reduces to the requirement that ${\partial}{\phi}/{\partial}\phi_o = 0$. 

An expression for the time dependence of the azimuth of a p-ellipse orbit is given by equation (C2) of S06, which can be written,

\begin{multline}
\tan \left( \frac{hmt}{2p^2} \sqrt{1-e^2} \right) = \\
\sqrt{\frac{1-e}{1+e}}
\left[ \tan \left(\frac{m \theta}{2}\right) -
\tan \left(\frac{m \theta_o}{2}\right) \right],
\end{multline}

\noindent where $h$ is the specific angular momentum of the orbit and $t$ is the time since impact. In general the parameters $p$ and $e$ will depend on the initial azimuth, $\phi_o$. However, in accordance with the previous section, the azimuthal dependence of $e$ and $p$ is weak at azimuths near $\phi = 3{\pi}/4$. Thus, when differentiating equation (21) with respect to $\phi_o$, we neglect the derivative of these parameters.

\begin{equation}
\frac{\partial}{\partial \phi_o}
\left[ \tan \left(\frac{m \theta}{2}\right) -
\tan \left(\frac{m \theta_o}{2}\right) \right] = 0.
\end{equation}

\noindent Combining this with the condition for the appearance of a (azimuthal) caustic we find,

\begin{equation}
\frac{\partial \phi}{\partial \phi_o} = 0 = 
\frac{\partial \theta}{\partial \theta_o} = 
\frac{\cos^2 \left(\frac{m \theta}{2}\right)}
{\cos^2 \left(\frac{m \theta_o}{2}\right)}.
\end{equation}

\noindent This is satisfied when,

\begin{equation}
\theta = \frac{\pi}{m} = \theta_o + \Delta \phi.
\end{equation}

\noindent For stars initially on the backbone this reduces to $\theta = {\pi/\sqrt{2}}$ and $\phi = 5{\pi}/4$, when $\delta = 0$. This value of $\theta$ shows that a star encounters the caustic at a fixed phase (outward moving) of its p-ellipse orbit. Since the tail is primarily made up of stars originating near the backbone the second equality implies that the azimuth of the tail caustic is close to $\phi = 5{\pi}/4$. At this level of approximation the shape of the tail is close to a straight line, while the caustic persists.

Like the caustic edges of collisional ring waves discussed in \citet[also \citealt{st10}]{ap96}, this tail caustic is an archetype. It is a generic consequence of any flyby disturbance strong enough to produce a clear tail.

Equation (24) only gives a leading edge caustic. Equation (23) can give others (e.g., a trailing caustic), and a less approximate treatment of the caustic condition could yield still more. The tail caustics generally end in an outer cusp. The inner base could either be another cusp, or a more complex caustic. Numerical images described below show that swallowtail caustics (see e.g., \citealt{ar86}) are common. 

The leading edge caustic is commonly seen in self-consistent numerical simulations of tails (e.g., \citealt{ge94}). Given that gas will pile up there, it is a likely site for star formation. This is especially true at the early stages of tail development when strong compression in the caustic, and as yet moderate stretching along the tail, may give optimal conditions.

\subsection{More caustics and swallowtails}

The backbone sector represents the basic tail dynamics, but the sectors $90^{\circ}$ away at initial azimuths $\phi_o = {\pi}/4$ and $5{\pi}/4$ tell a different story. According to equation (11), at these azimuths $m\theta_o = \pi$ and $\psi = 0$. Equation (14) yields,

\begin{equation}
\frac{p}{q} = \left( 1-e \right)^{\frac{1}{2} + \delta},
\end{equation}

\noindent The analog of equation (18) in this case is,

\begin{equation}
\frac{r_{ap}}{q} = 
\left(\frac{1-A'}{1-e} \right)
^{\left(\frac{1}{2} + \delta\right)}
\approx \left(\frac{1-A'}{1-A'} \right)
^{\left(\frac{1}{2} + \delta\right)}
= 1.
\end{equation}

\noindent Thus, the orbit size is decreased by the impulse along these sectors, and the initial radius is the apoapse. The radial velocity impulse is zero along these sectors and the azimuthal impulse is opposite the star's rotation. Stars along these and nearby sectors will fall inward and pile up at periapse, forming caustics in the inner disc. Alternately, at large radii, where the perturbation is strongest, they can contribute to the intersecting stellar streams that make swallowtail caustics at the base of the tidal tail.

The sectors around initial azimuths of $\phi_o = 0$ and $\pi$ help complete the story. They are pulled outward, away from the nucleus by the tidal perturbation. That is, the disturbance provides an outward radial velocity, and small or zero azimuthal velocity perturbation. This perturbation is like that in the disc of a colliding ring galaxy, except $180^{\circ}$ out of phase (pulled outwards instead of inwards). Stars in these sectors will first move outward, and then inward. In the outer disc they can contribute another star stream to the swallowtail caustics at the base of the tidal tails. Swallowtails have five intersecting streams, in contrast to the three in ring wave caustics. 

Stars in sectors around initial azimuths of $\phi_o = {\pi}/2$ and $3{\pi}/2$ will be pulled inward by the tidal perturbation even more like ring galaxies. However, this pull is relatively weak, and at best only generates a weak wave in the outer disc. 

\subsection{Ocular waves}

It is worth briefly examining the approximate radial caustic condition before leaving the topic of caustic waves. This condition is ${\partial}r/{\partial}q = 0$, i.e., the complete compression of an initially finite range of radii. Using equation (2) for $p$ in equation (1), differentiating with respect to $q$, and then simplifying the result we obtain, 

\begin{multline}
\frac{\partial r}{\partial q} = \left[
\frac{1 + e\cos \left(m \theta_o\right)}
{1 + e\cos \left(m \theta\right)}\right]
^{\frac{1}{2}+\delta} \times\\
\left[1 + \left( {\frac{1}{2}+\delta} \right) q
\frac{\partial e}{\partial q}
\left[ \frac{\cos \left(m \theta_o\right) - \cos \left(m \theta \right)}
{\left({1 + e\cos \left(m \theta\right)}\right) \left({1 + e\cos \left(m \theta_o\right)}\right)}
\right]
\right].
\end{multline}

\noindent As long as $e < 1$, the caustic condition can only be satisfied when the second term on the right-hand-side (in the outer square brackets) is zero.  Thus, the approximate caustic condition is,

\begin{multline}
0 =  1 + \\
 \left( {\frac{1}{2}+\delta} \right) q
\frac{\partial e}{\partial q}
\left[ \frac{\cos \left(m \theta_o\right) - \cos \left(m \theta \right)}
{\left({1 + e\cos \left(m \theta\right)}\right) \left({1 + e\cos \left(m \theta_o\right)}\right)}
\right].
\end{multline}

\noindent Note that with $e$ given by equation (12) (or equation (16)), with substitutions from equations (7) and (9), the value of the factor $q({\partial}e/{\partial}q)$ will be within a factor of a few of $e$, for the power-law potentials considered here. Given that the factors in front of the term in square brackets are less than or about unity, we want to minimize the negative value of the term within the square brackets to satisfy the equation (or make the right-hand-side of the equation negative, indicating an orbit between two caustics). Evidently, this is achieved near $m\theta_o = \pi$ and $m\theta = 2\pi$. Equation (12) then implies that $\phi_o = \pi/4$ or $5\pi/4$ (perpendicular to the backbone). Then, $\phi$ is given by, $\Delta \phi = \Delta \theta$ (equation (24).) Thus, with the values of the other variables just given we find that $\phi = \pi/4 + \pi (1/2 + \delta) m^2$ or $\phi = 5\pi/4 + \pi (1/2 + \delta) m^2$. In the logarithmic (Keplerian) potential with $\delta = 0\ (1/2)$, then $\phi = 3\pi/4$ or $7\pi/4$ (in both cases) to satisfy the caustic condition.

As noted in the previous subsection, stars near these initial azimuths lose angular momentum from the impulse and fall inward. Equation (28) suggests that their pileup at periapse generates a radial caustic if $e$ and its gradient are sufficiently large. I.e., when the impulse is sufficiently strong. These caustics appear at azimuths orthogonal to the angular caustic in the tail. These facts suggest the identification of the radial caustics as ocular waves of the type described by \citet{el91}. The present analysis agrees with numerical results in those references in several respects, including the association of ocular waves with nascent tails, the orthogonal orientation of these two features, and the existence of a threshold perturbation amplitude to generate the oculars. Though again very approximate, the present analysis suggests that the ocular waves are another caustic archetype.

In the previous subsection we also noted the presence of a radial oscillation component if the orbits of stars at initial azimuths of $\phi_o = 0$ (or $\pi$) and $\phi_o = \pi/2$ (or 3$\pi$/2). In these cases, equation (11) yields values that make $\cos(m\theta_o) = 0$. Then, inspection shows that while the last term in equation (28) can easily be negative, it is unlikely to reach a value of one and satisfy the caustic condition. Undoubtedly, there will be radial compressions of the stars near these initial azimuths, but strong radial caustics are unlikely. We also note that slowly developing spiral caustics of the type considered in \citet{st11} are not captured by the rough analysis here, though they are related.

\section{Graphical realizations of the analytic results}
\subsection{Illustrating the analytical discs and tails}

The simple analytic model of the previous section yields many powerful insights on disc and tail structure. However, the full tidal structure of this model can only be revealed by looking at a large sample of disc orbits described by the analytic equations. This is best done graphically. Some results are shown in Figures 4-8. 

Each of these figures shows the positions of 38,100 stars initially distributed uniformly in the disc, and assumed to have circular velocities in centrifugal equilibrium with a power-law gravitational potential. Subsequently, the orbits of these stars were computed according to the analytic solutions of the previous section following an impulsive disturbance of specified magnitude (and with equation (16) used for the eccentricity). In addition the azimuths of the stars as a function of time after the impulse on p-ellipse orbits are given by a nonlinear, multi-valued function of time (equation (C2) of S06). Rather than find the appropriate solutions to this equation for every time, we numerically integrated the equation for the azimuthal velocity for each star to each desired time value. 

\begin{figure}
\centerline{
\includegraphics[scale=0.36]{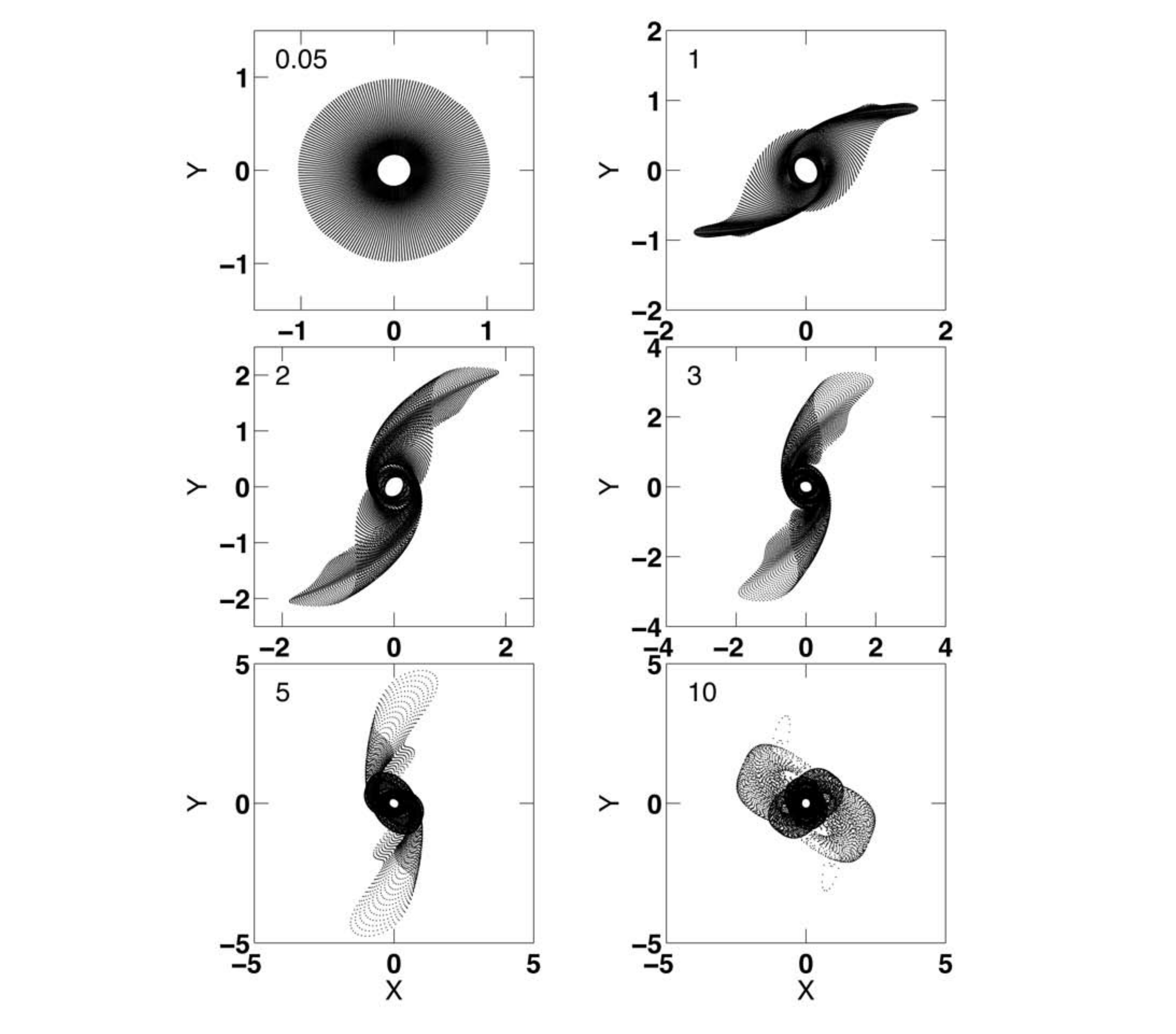}}
\caption{Six snapshots of the evolution of an analytic model disc in a logarithmic ($\delta = 0$) potential after a tidal impulse at time $t = 0$. The amplitude of the impulse is $A' = 0.80$ (at $q = 1.0$), the initial disc radius in dimensionless units). The orbital period at the initial disc edge is $2\pi$ dimensionless time units. Note the different scales of the different panels. See text for details. }
\end{figure}

\begin{figure}
\centerline{
\includegraphics[scale=0.4]{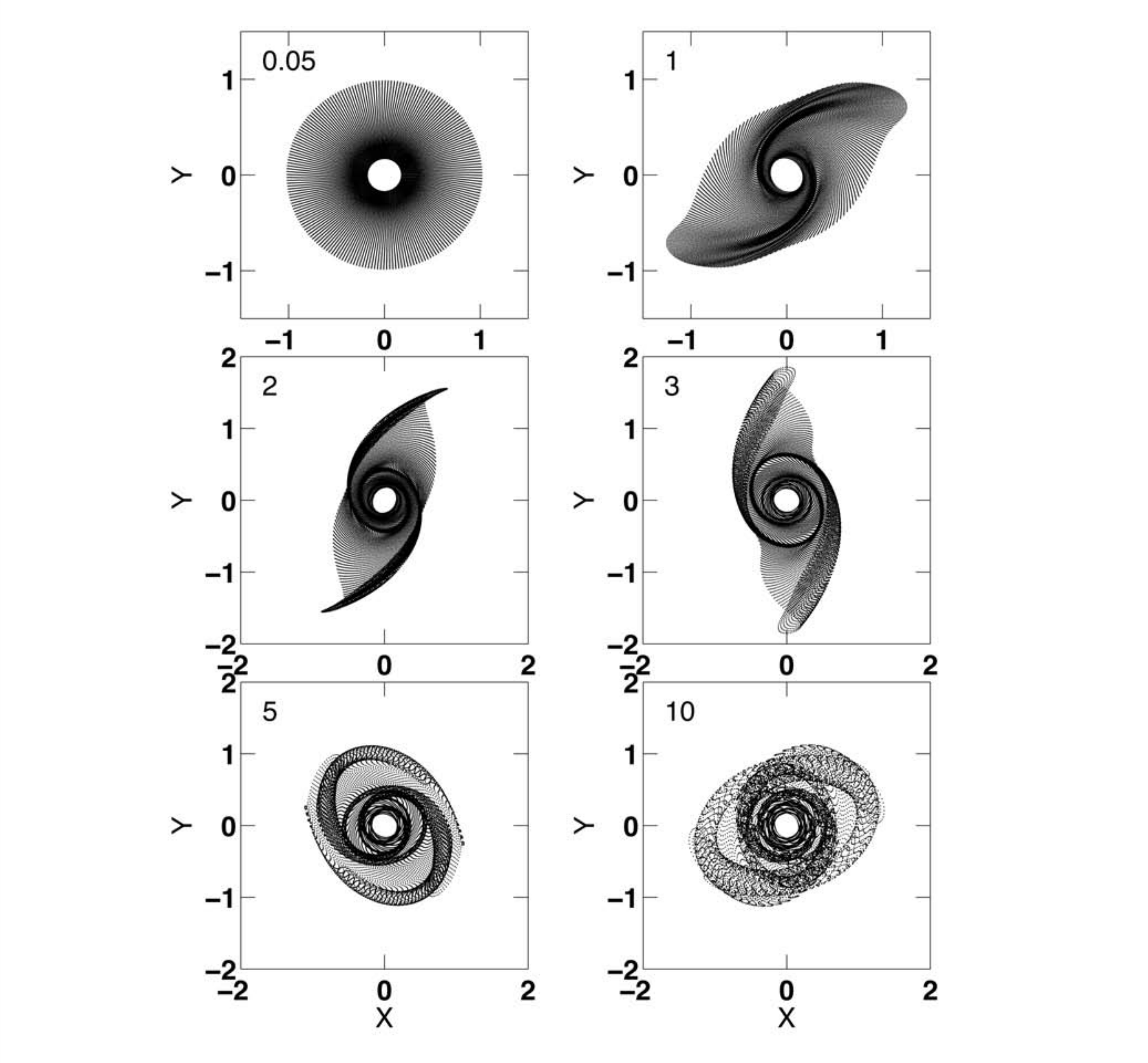}}
\caption{Same as Figure 4, but with a tidal impulse amplitude of $A' = 0.50$ (at $q = 1.0$).}
\end{figure} 

\begin{figure}
\centerline{
\includegraphics[scale=0.4]{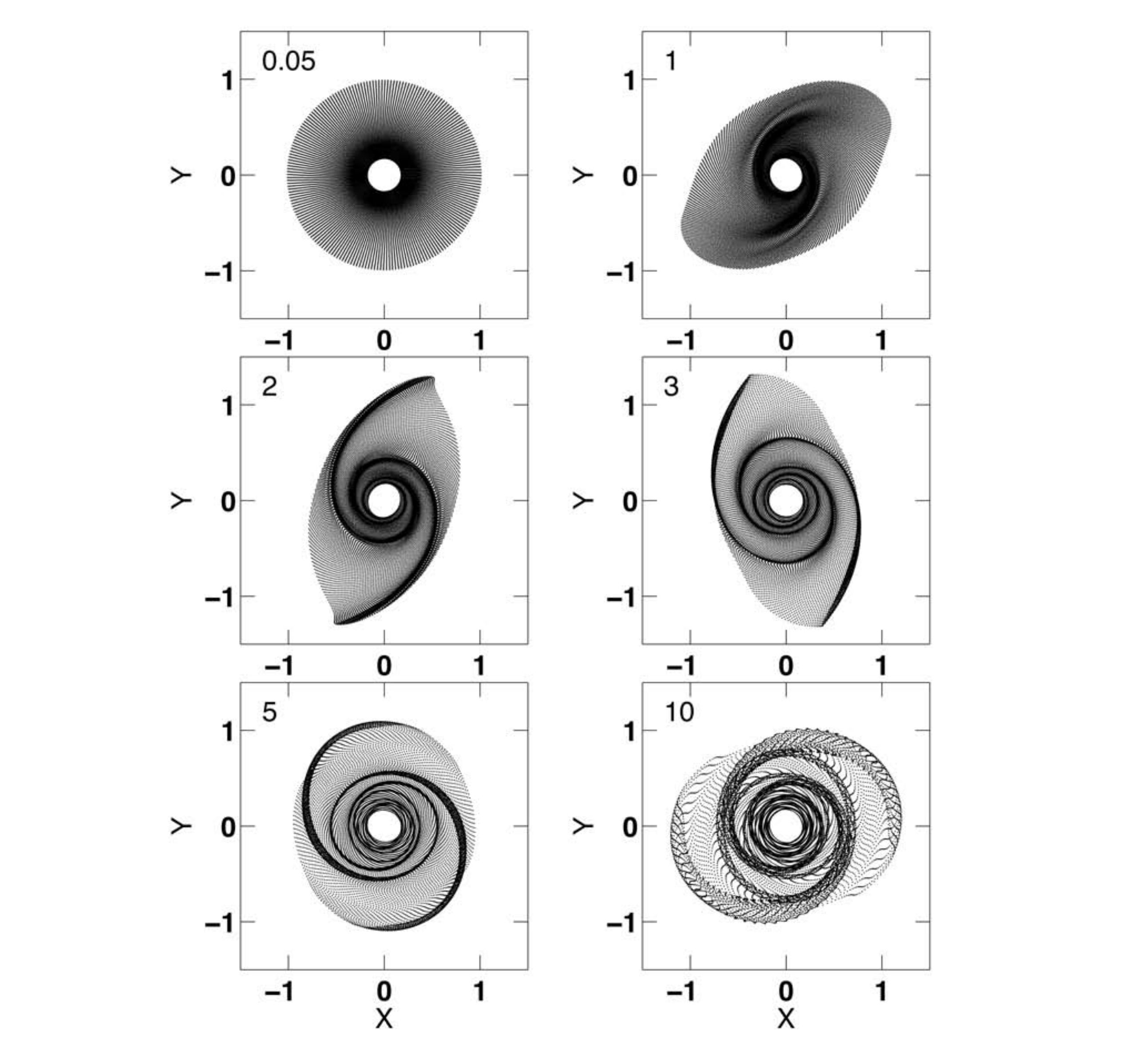}}
\caption{Same as Figure 4, but with a tidal impulse amplitude of $A' = 0.30$ (at $q = 1.0$). At this amplitude, or lower, the tails appear more like internal disc spirals.}
\end{figure}

In this section we focus on the results for the logarithmic and Keplerian potentials, which bound the range of physically interesting flat-to-declining rotation curve slopes. Specifically, Figures 4-6 each show six snapshots of the post-collision disc in the logarithmic potential, with different perturbation amplitudes in each case. 

In these models the units are chosen such that the radius of the disc is 1.0, and $GM(R)$, the gravitational constant times the mass contained within the disc, is also set to 1.0. Then, the circular velocity at the outer radius is 1.0, and the orbital period is 2$\pi$. In this paper we use only these dimensionless units, but a consistent set of scaling factors is, for example, $R = 10$ kpc, $v(R) = 250$ km s$^{-1}$, $M(R) = 1.5 \times 10^{11}$ M$_{\odot}$, and $P(R) = 250$ Myr. 

Figure 4 shows the highest amplitude case, with $A' = 0.8$ for particles at the outer edge of the disc. The first 4 panels of this figure show the steady growth of the tail. Tail particles extend to a radius of nearly 5 times the radius of the initial disc, despite the fact that the disturbance is impulsive. This value is in good accord with that predicted by equation (18) (as it should since both are realizations of the same approximation). The final panel shows the orbital excursions of groups of stars into the halo at a late time when the orbital phases at different initial radii have drifted too far to allow the formation of coherent structures. Although tail stars are perturbed to very radial orbits, they cannot form a recognizable tail on their second passage out into the halo. The lifetime of the tail is roughly the orbital period of the particles flung to the greatest distances. 

Similar results are seen in the milder case, $A' = 0.5$, shown in Figure 5. Here again the maximum tail length is well predicted by equation (18). The formation of azimuthal caustics on the leading edges of the tails is also visible in this figure. They form at about the time shown in the second panel, near the expected azimuths ($\pi/4$ and $5\pi/4$, see Sec. 2.7).

Caustic edges are even more visible in the still milder case, $A' = 0.3$,  shown in Figure 6. Here the tails look more like spiral waves in an oval disc than tidal tails. All three of these figures have some general characteristics in common. All are bilaterally symmetric as they must be given the symmetry of the simple tidal perturbation of equations (5) - (7). 

The tidal tails, and especially their compressed leading edges, are relatively straight in this approximation. Figures 7 and 8 show that they are not as straight in the Keplerian potential. These figures also confirm the point emphasized by \citet{du04}, that with other parameters equal, Keplerian tails are much longer than those of logarithmic potentials. Quantitatively, the maximum radial extents are in good agreement with equation (18), with $\delta = 1/2$.  At some times the Keplerian tails look unrealistically broad compared to typical optical images. However, galaxies do not generally have rotation curves that decline as rapidly as the Keplerian. 

\begin{figure}
\centerline{
\includegraphics[scale=0.4]{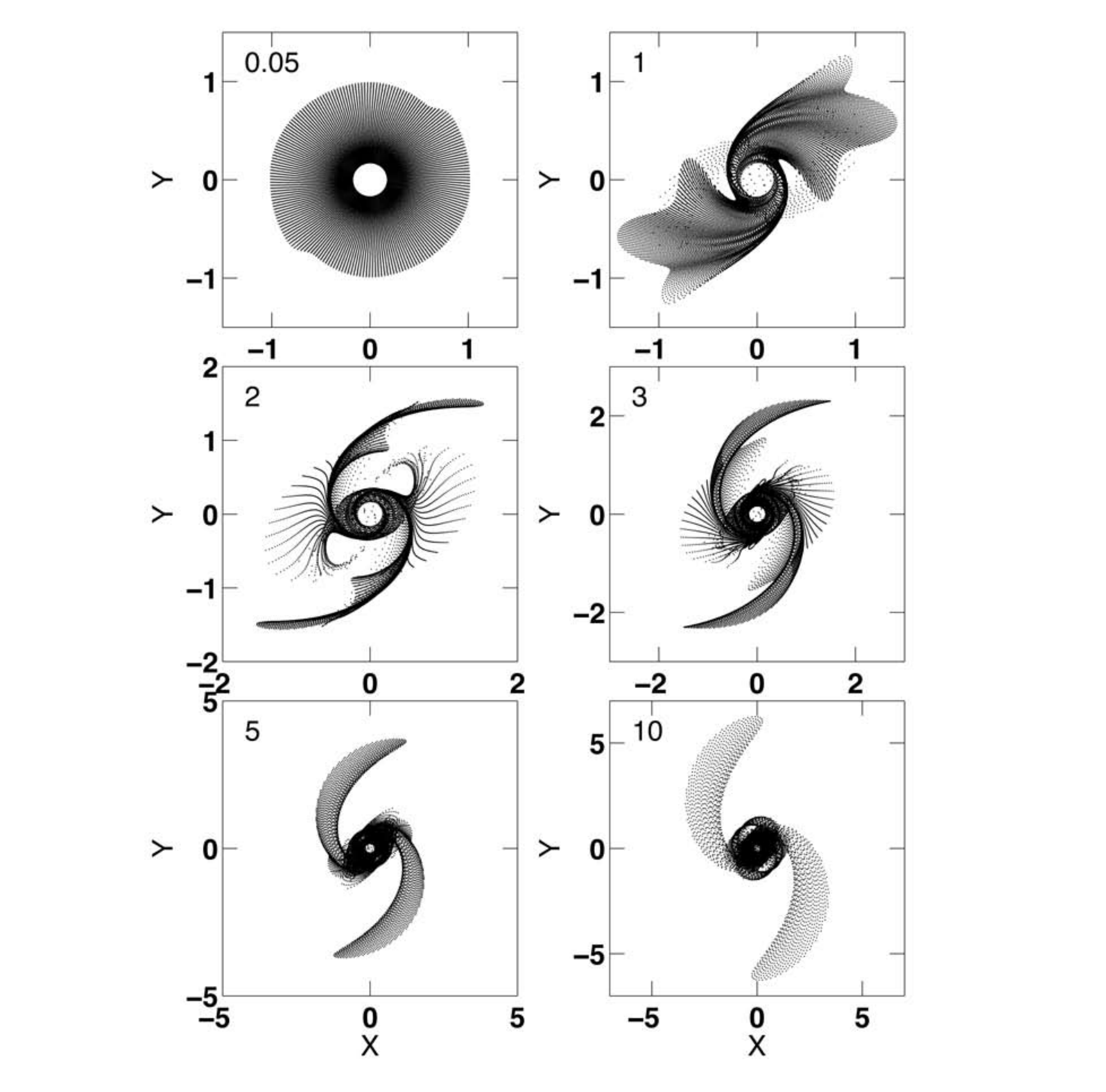}}
\caption{Like Figure 4, but with a Keplerian ($\delta = 0.5$) potential, and a tidal impulse amplitude of $A' = 0.35$ (at $q = 1.0$). Note that the appearance of the thin filaments in the middle panels is an artifact of the initial conditions.}
\end{figure}

\begin{figure}
\centerline{
\includegraphics[scale=0.4]{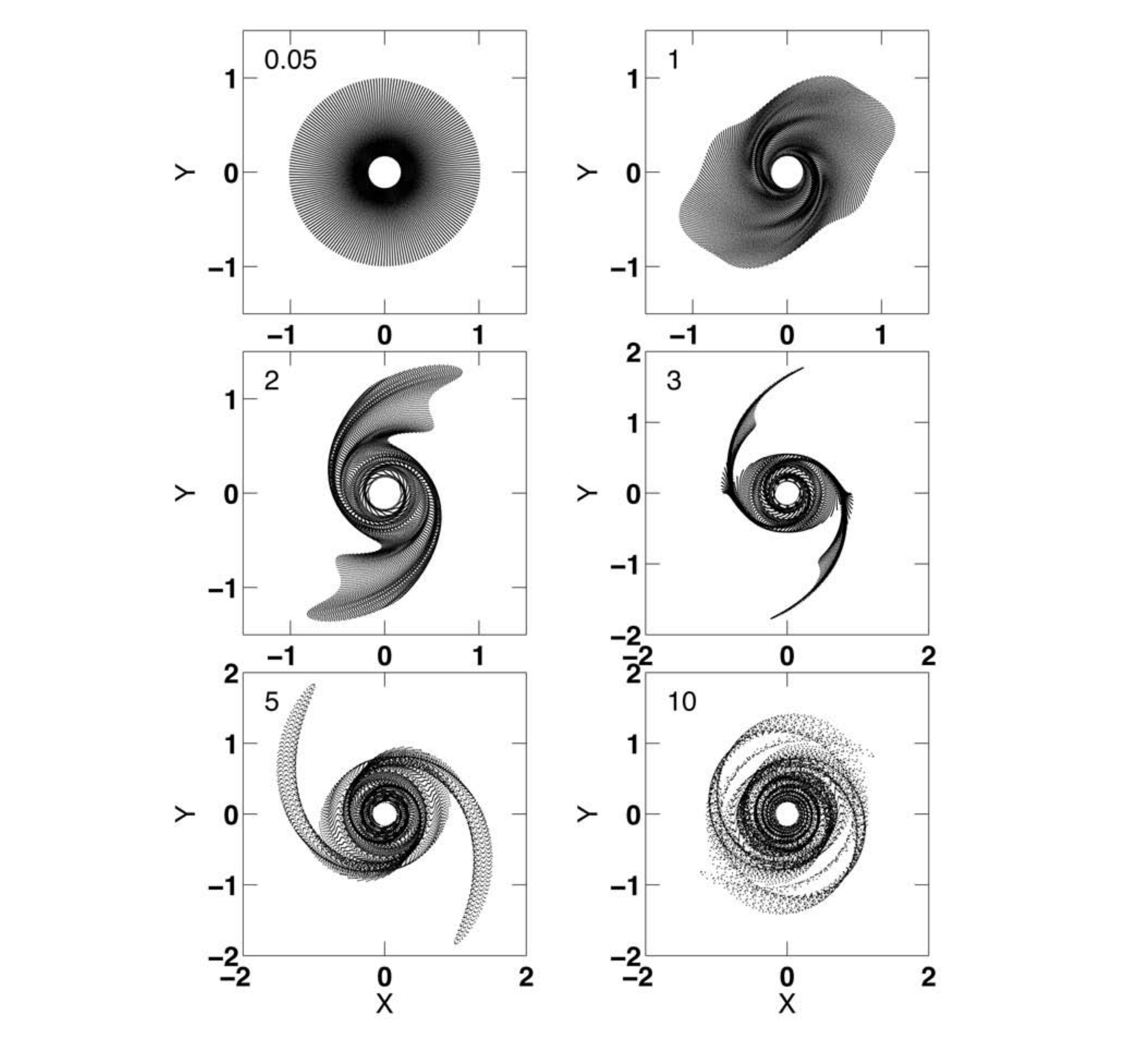}}
\caption{Same as Figure 7 (Keplerian potential), but with a tidal impulse amplitude of $A' = 0.20$ (at $q = 1.0$).}
\end{figure}

This effect is further illustrated in Figure 9, which shows the tails that develop after 5.0 time units in six model discs differing only in the exponent of their power-law potentials. (Compare the second to last panels in Figs. 6 and 7.) The upper left panel shows the case of the logarithmic potential ($\delta = 0$), and the lower right panel show the Keplerian potential ($\delta = 0.5$). The other panels have intermediate values of the exponent $\delta$ and show a continuous increase in the tail length toward the Keplerian case. Otherwise the tails are qualitatively similar across the range of potentials.

\begin{figure}
\centerline{
\includegraphics[scale=0.35]{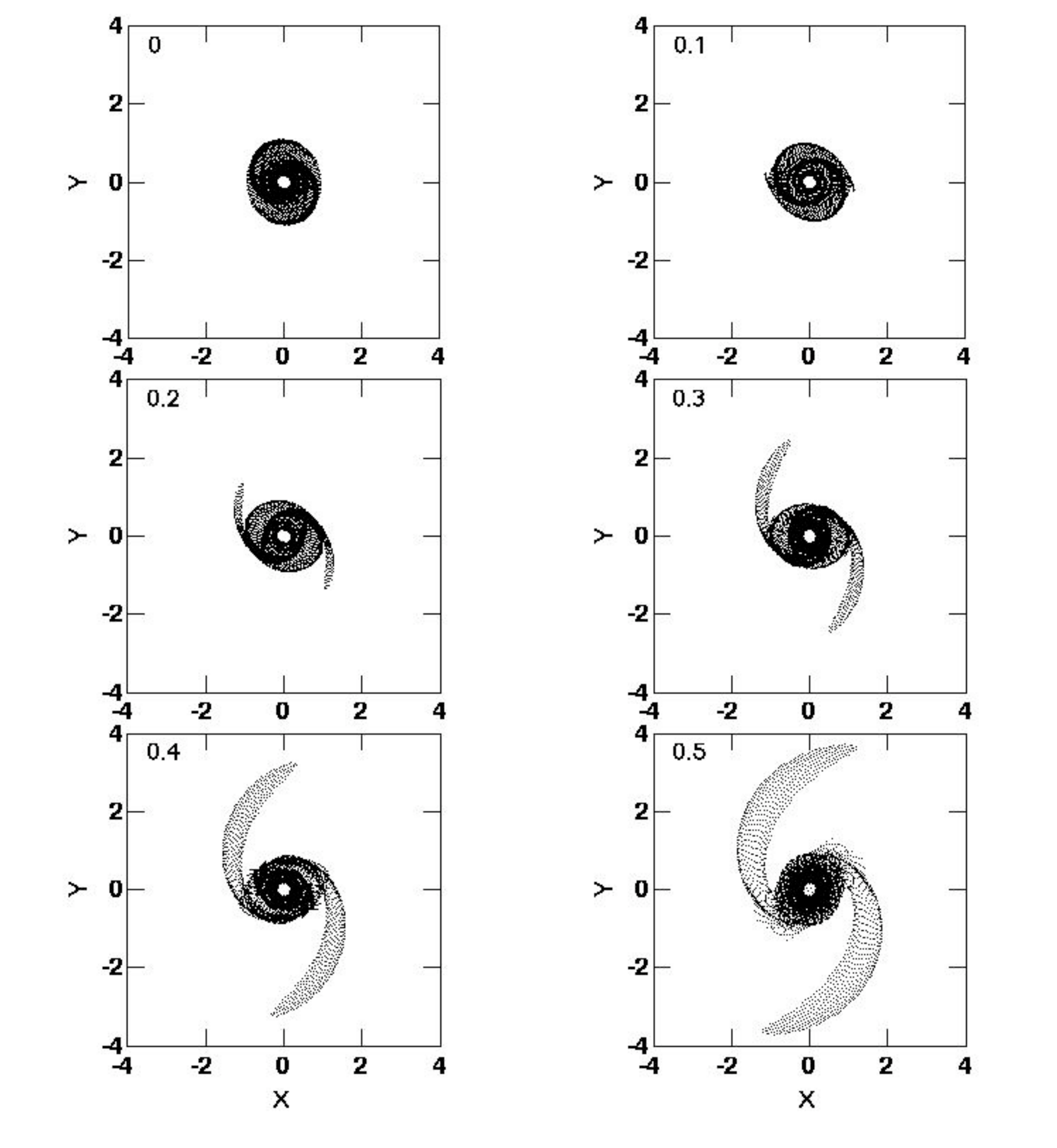}}
\caption{Analytic model discs formed from a tidal perturbation of amplitude $A' = 0.35$ (at $q = 1.0$). Each panel shows a model with a different value of the potential exponent $\delta$, which is given in the upper left corner. In all cases the model is shown at a time of 5.0 units.}
\end{figure}

Figures 4-8 provide a large-scale view of the discs and tails, while we give one example of detailed structure within the disc in Figure 10. This figure shows results of a $A' = 0.6$ amplitude encounter, with a potential intermediate between the logarithmic and the Keplerian ($\delta = 0.2$). The model produces substantial tails. The inner disc is essentially an ocular structure (centred at about the azimuths predicted in Section 2.9), with strong internal spirals. Associated with the ocular and spirals the disc also produces strong swallowtail (or higher order) caustics. E.g., paired ones with their strongest cusps located at about the 1 and 7 o'clock positions in the figure (or about PA $-30^{\circ}$ and $150^{\circ}$). Presumably dissipative gas piles up in the converging flows that define the densest regions of the swallowtails, so these are good candidate sites for induced star formation (see e.g., \citealt{re09}, \citealt{re11}). The outer caustic regions could be the locus of Ôhinge clumpsÕ discussed below. 

Some caveats should be noted, however. First, the extreme symmetry of these structures is the effect of the assumed pure tidal perturbation, and is not realistic. Secondly, such structures are not common at this level of approximation. Their production requires high amplitudes, very high amplitudes in the case of the logarithmic potential. These caveats are relaxed in the case of nonlinear, asymmetric perturbations considered in Sec. 5. However, first we must consider another extension to the case of simple tidal perturbations.

\begin{figure}
\centerline{
\includegraphics[scale=0.4]{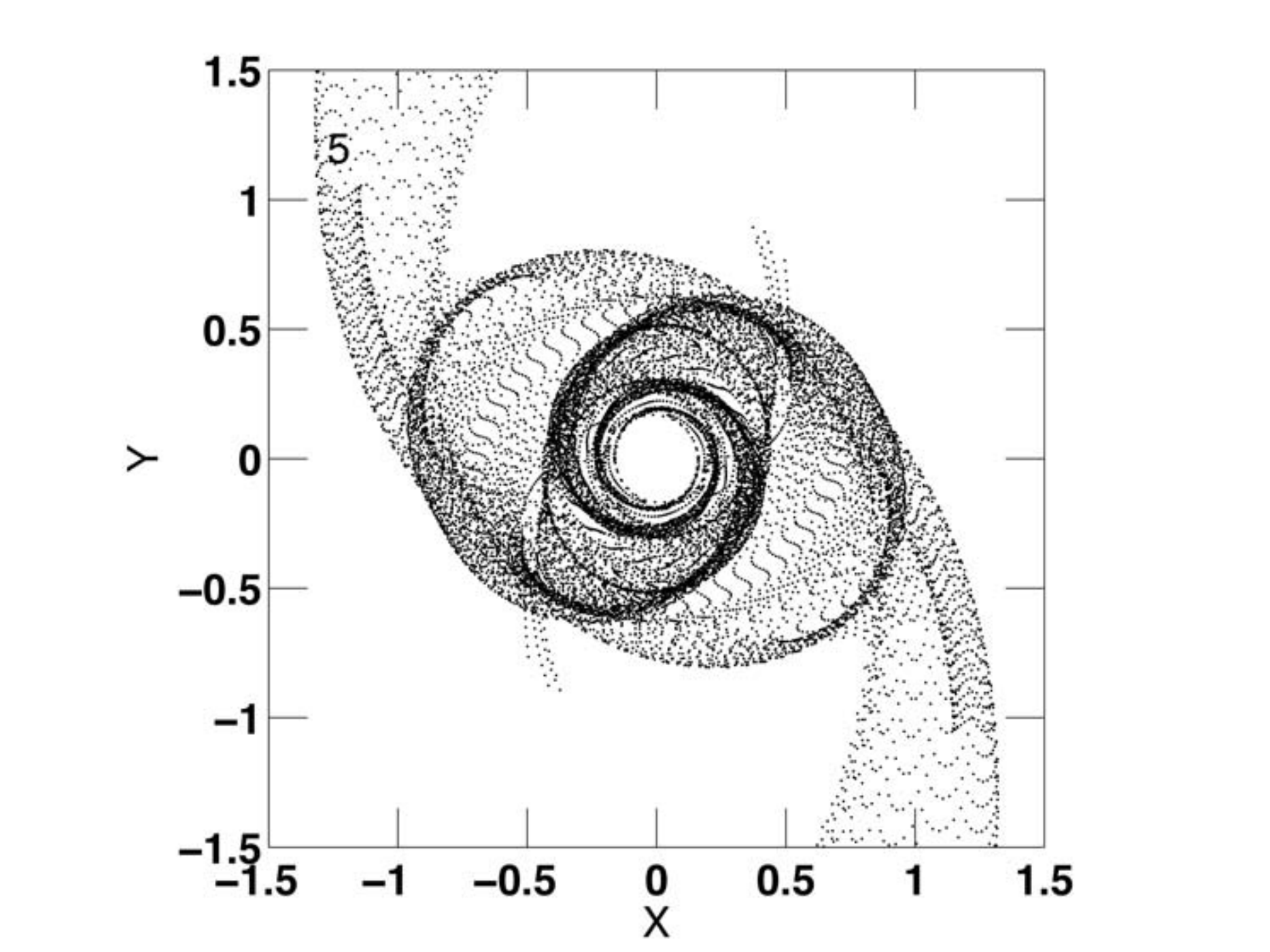}}
\caption{Snapshot enlargement of a model disc with a potential intermediate between the logarithmic and Keplerian ($\delta = 0.2$) at time $t = 3.0$ units, and an amplitude of $A' = 0.60$ (at $q = 1.0$). Paired, higher order caustics (e.g., swallowtails) are evident in the outer disc near the base of the tails. }
\end{figure}

\subsection{Brief comparisons to numerical discs and tails}

Figures 1-3 and the discussion concerning them provide a basic view of the accuracy of the analytic models, but much more analysis would be required for a complete evaluation. That is beyond the scope of this paper, but following the theme of this section, we will pursue the issue somewhat further via another graphical comparison. This is provided by Figure 11, which was created for comparison to Figure 5. All model parameter values are the same in both figures, including the tidal impulse amplitude. The difference is that following the impulsive perturbation to the initially circular orbits, the particle trajectories were computed numerically for Figure 11. Of course the result is not the same as a fully self-consistent, N-body simulation, but is a step in that direction, with self-consistent orbits calculated in the fixed potential. 

The results in Figure 11 clearly differ in some ways from those of Figure 5. The tails are somewhat longer at comparable times (note the slightly different scales of the two figures), and longer lived in the numerical model. In terms of these properties, analytic models with larger tidal amplitudes (e.g., $A' = 0. 6$) provide a better fit to the numerical results. However, given that the analytic model is formally accurate only to order $e$, the overall agreement is very good. 

Another clear difference is between the thicknesses of the tails in the two figures. A more precise description would be that the dense, but thin tails, in Figure 5 have not separated from the parent disc, which is more elongated than that of Figure 11. We can understand this with the help of Figure 1, where in the 4 trajectories at the bottom of the figure the analytic curves usually lie significantly above the numerical curves, i.e., at significantly larger radii. Nonetheless, these differences are again well within the expected limits of the approximation. 

\begin{figure}
\centerline{
\includegraphics[scale=0.35]{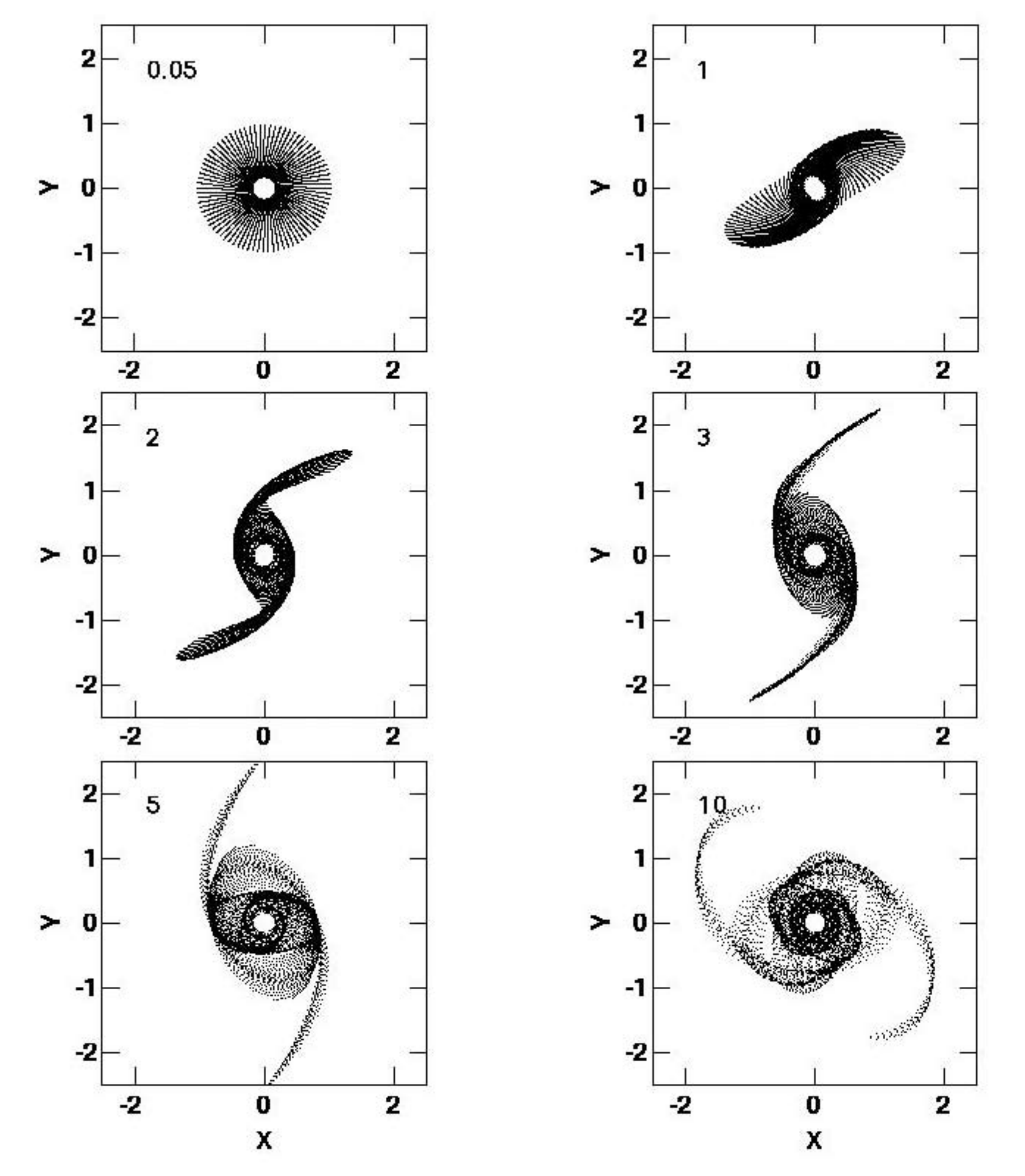}}
\caption{Same as Figure 5, except particle trajectories are computed numerically. See discussion in text.}
\end{figure}

\section{Effects of two impulses}

Calculations of two impulses separated by a finite time interval can provide the simplest model for a prolonged disturbance, and thus, the simplest representation of the difference between prograde and retrograde encounters between disc galaxies. In this section we will explore analytic approximations to this case. Most of the relevant equations, and some limits are described in the appendix.

\subsection{Prograde versus retrograde interactions}

The equations derived in the appendix show that, in optimal circumstances, the apoapse radius of a stellar orbit can grow geometrically with successive impulses. What is required for this optimal amplification? Ultimately it is due to two distinct physical effects, which correspond to the numerator and denominator of the term in square brackets in equation (45). The first effect, codified in the numerator of the term in equation (45), is the increase in angular momentum resulting from the prograde perturbation (see eqs. (6) and (38)). More specifically, it concerns the region where stars receive the maximum increase in angular momentum. In the first impulse this was along the azimuthal line called the backbone in Section 2.5. Clearly, if stars on the backbone also receive the largest angular momentum increase in the second impulse their amplitude and apoapse radius will be relatively large. Such stars must lie on both the first or second Ôbackbones,Õ i.e., at azimuths of $-\pi/2$ or $3\pi/4$ relative to the closest approach point at both impacts. 

This is most easily achieved if these stars rotate at the same average angular frequency as the companion (or equivalent angular frequency in the case of hits by two different companions). Since angular frequency in the disc generally varies with radius, generally at most one spot of the original backbone (call it a particular ÔvertebraÕ) will satisfy this condition, and be found on both backbones. 

For most stars, their relative azimuth at the second impact will be different from the first. They will not get the maximum boost from both, and in some cases the angular momentum impulses may be of opposite signs. This is the interference between the two impulses noted above. In the case of a retrograde interaction, it is practically impossible for a star on the original backbone to find itself on the second. (Unless, of course, the companion and the disc stars have combined rotations of more than $2\pi$ radians between the two impulses.) The retrograde case also generates an interference pattern of velocity impulses on the disc, but one in which the original winners (in terms of angular momentum gains), are much less likely to win again. 

The second factor, for determining the amplitude of a stellar orbit after the second impulse, is given by the denominator of the bracketed term of equation (45). Physically, the value of this term is determined by the phase along the p-ellipse orbit at the time of the second impact. I.e., if the star is at apoapse it will be most strongly affected. The optimal effect, described by equations (47) and (51), occurs for stars that lie on both backbones and happen to be at apoapse at the time of the second impact. 

To get the maximal input from this second factor, the azimuthal phases of the p-ellipse immediately after the first impulse and just before the second, must be given by, 

\begin{equation}
\theta_{I2} = \theta_{I1} + c\Delta \phi = \frac{\pi}{m},
\end{equation}

\noindent Using equation (12) we solve for the total azimuthal phase change,

\begin{equation}
\Delta \phi = \frac{\pi}{m} - 
\left( 2\phi_o +\frac{\pi}{2} + \psi \right).
\mbox{          (Modulo 2$\pi$.)}
\end{equation}

\noindent The two amplitude effects are not directly connected, so such a conjunction between them would appear to be a coincidence. However, there is a bit more to it than that. The second effect requires a significant time between impulses, while the first effect requires co-rotation between a region on the backbone and the companion. 

If the two impulses represent a prolonged encounter, then the coincidence of the two effects does not seem unlikely.  At the least, the continued angular momentum input into a particular region on the initial backbone, lasting through half an orbital period of a kind of average p-ellipse orbit in that region, seems reasonably likely. It is certainly in accord with many published simulations which show that the longest tails are produced by prolonged prograde encounters, and is the simplest analytic representations of such encounters (see also \citealt{do10}, and \citealt{ne99}). The present treatment shows clearly the effects of resonance in the azimuthal frequency.

\subsection{Conjectures on mechanisms of tidal dwarf formation}

It is likely that there are several specific mechanisms for producing dwarf galaxies in tidal tails. One of these may be based on the azimuthal pileup discussed in Section 2. However, a number of the more substantial tidal dwarfs seem to lie at or near the end of tails. This requires a mechanism of radial pileup. Nothing like that was evident in the impulsive tails described in the previous sections. On the other hand, the phenomenon described in the previous subsection of a `resonant vertebra' along the initial backbone of the tail, which receives the strongest azimuthal velocity input in the disc from a prolonged encounter, could lead to such an effect. 

With greater angular momentum than surrounding regions, stars in the resonant vertebra region could overtake others in the tail, which were initially located at larger radii, leading to a density enhancement. In the case of gas clouds this could lead to an actual pileup or shock wave. Both effects could lead to local gravitational instability and the formation of a self-gravitating tidal dwarf. Regions interior to the resonant vertebra would be tidally stretched more strongly than in single impulse tails, and as a result, would be less likely to produce large star clusters.  However, there could be a competition with the effects of azimuthal pileup, though this is likely to occur at an earlier time than the radial pileup. These considerations could explain the observed tendency for tails with relatively large dwarfs at their ends to have few clusters along their length, and vice versa (\citealt{kn03}, \citealt{mu11}, \citealt{ka12}). 

They could also help explain the rarity of large dwarfs at the end of tails. If the resonant vertebra is located at the outer edge of the initial disc (or beyond), there will be no strong pileup. If the resonant vertebra is located deep within the initial disc, then the pileup will more likely occur within the disc rather than well out in the tail. This might instead play a role in the phenomenon of Ôhinge clumpsÕ (as defined in \citealt{ha09}, \citealt{sm10a}, \citealt{sm10b}). 

In any case, the discussion in these last two subsections illustrates the exploratory power of the two-impulse, p-ellipse model for disturbed stellar orbits, and their collective behavior.

\subsection{Graphical realizations with multiple hits}

Two sample realizations of tail evolution with multiple hits are shown in Figures 12 and 13. In both these models there are three successive (prograde) tidal impulses, each with an amplitude of $A = 0.35$. The formalism for two hits given in the preceding sections is straightforwardly extended for any number of impulses, and readily implemented in a computer script. The choice of three impulses here is arbitrary. It is sufficient to illustrate the effect of multiple hits and compare to the formulae above, but not as complex as a simulation of a very prolonged encounter with numerous impulses. 

In the case shown in Figure 12 the effective azimuthal frequency of the companion relative to that of a disc star is about $3.7q$. This ratio is equal to one at an initial disc radius of $q = 0.27$. The resonant vertebrae of the backbone should be located at about that radius, rather deep inside the disc. The azimuths of the three impacts are $0^{\circ}, 70^{\circ}$ and $140^{\circ}$. 

\begin{figure}
\centerline{
\includegraphics[scale=0.36]{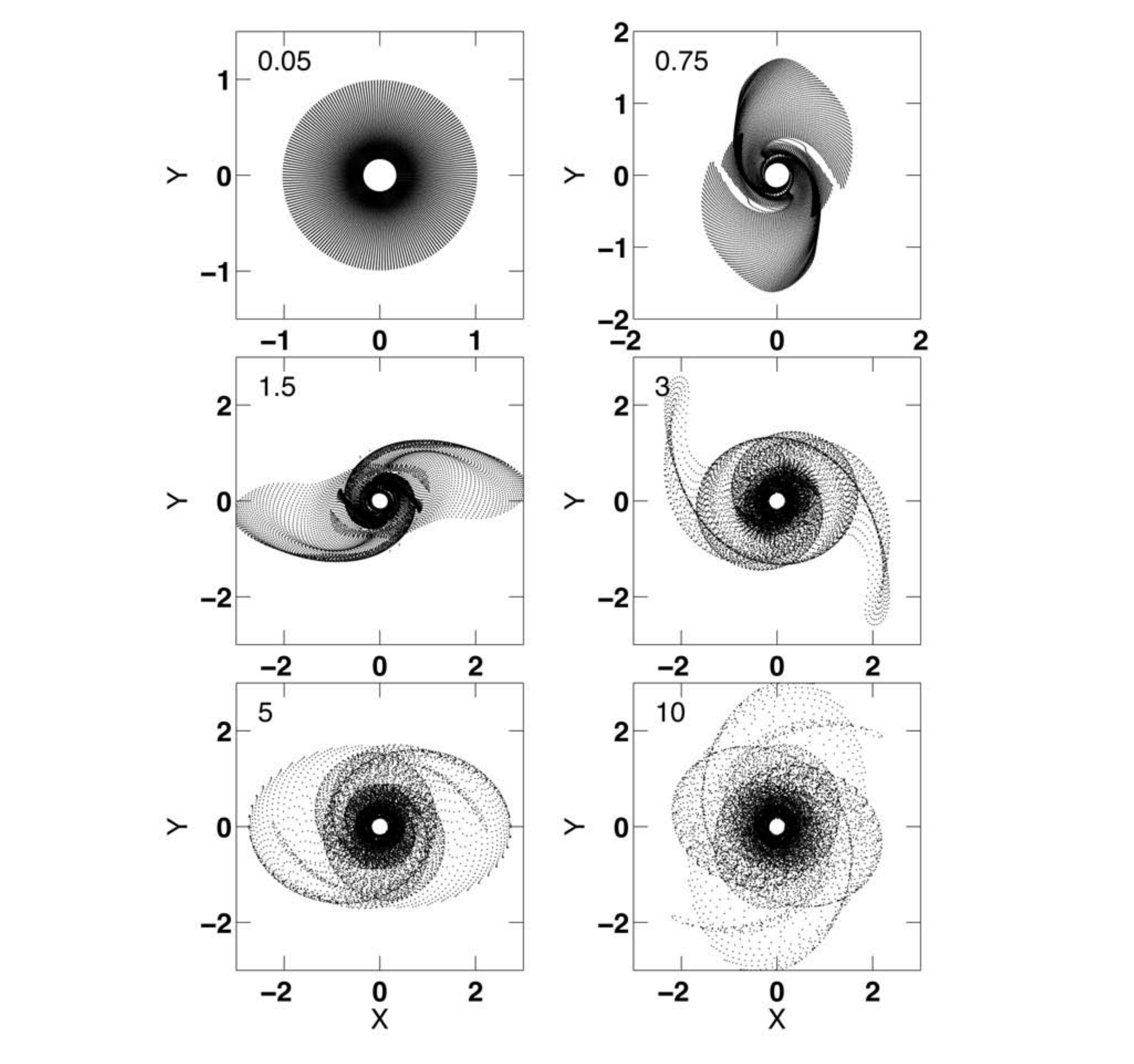}}
\caption{Snapshots of the evolution of a disc experiencing three tidal impulses, each of amplitude $A' = 0.35$ (at $q = 1.0$). The impulses occur at azimuths of about $0^{\circ}, 70^{\circ}$ and $140^{\circ}$, and at times of 0.05, 0.75 and 1.5 units.}
\end{figure}

\begin{figure}
\centerline{
\includegraphics[scale=0.4]{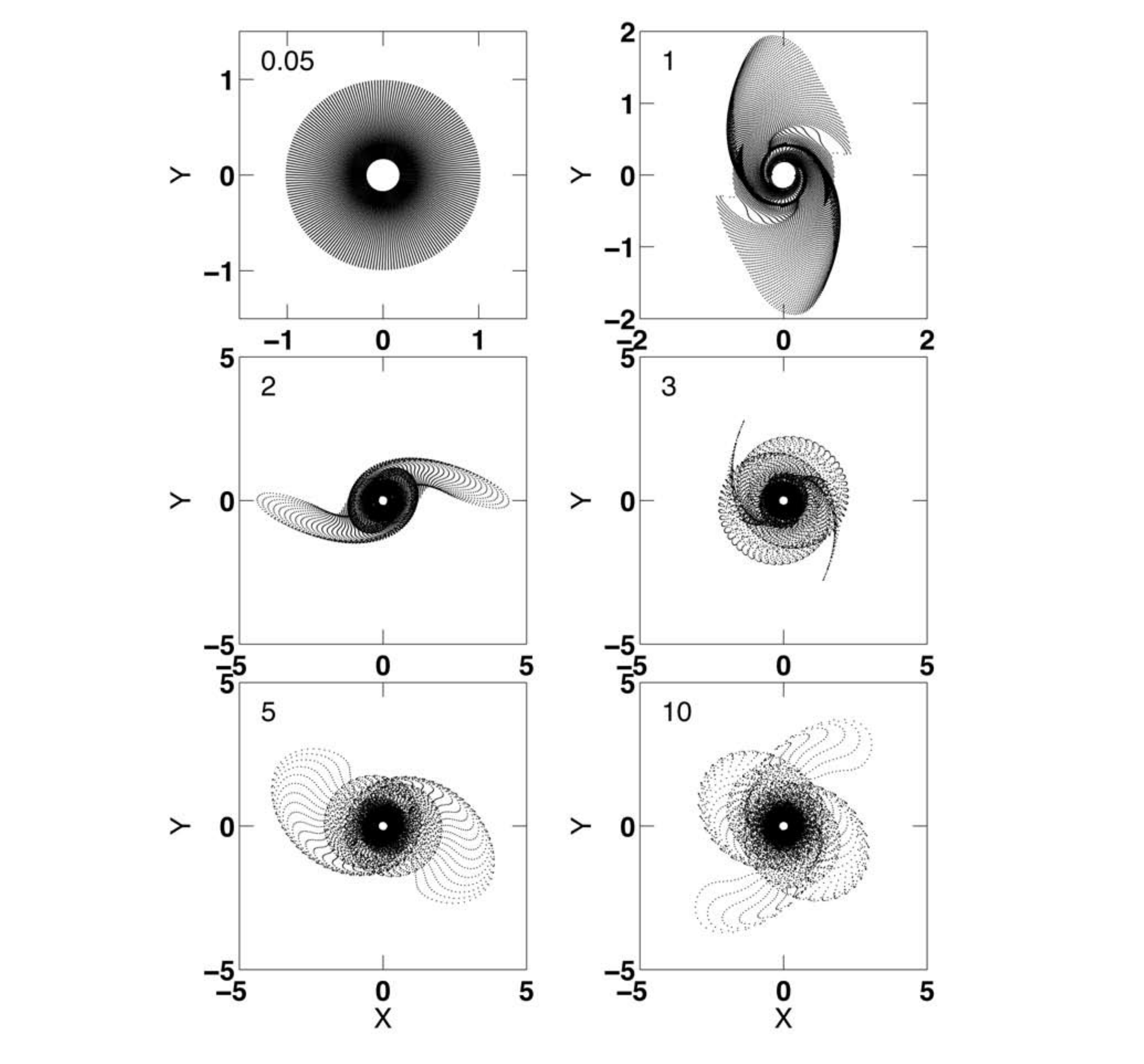}}
\caption{Same as Figure 10, except that the impulses occur at azimuths of about $0^{\circ}, 57^{\circ}$ and $114^{\circ}$, and at times of 0.05, 1.0 and 2.0 units.}
\end{figure}

For the case shown in Figure 13 the effective azimuthal frequency of the companion relative to a disc star is q. Clearly, this ratio is equal to unity at a disc radius of $q = 1.0$, at the outer edge of the disc. In this case, the azimuths of the three impacts are $0^{\circ}, 57^{\circ}$ and $114^{\circ}$.

Equation (47) suggests that the maximum amplitude in these cases is about 0.9. The approximate equation (51) suggests that the maximum obtainable apoapse radius is about $1.7^2 = 2.9$ times the initial radius. The better approximation of equation (50) suggests that this factor is more like 4.0.  In Figure 12 the tail does not stretch out to such a large factor relative to the initial disc radius. The maximum radius in this case is somewhat larger than 3. This is not surprising given that the strongest orbital resonances are with stars located well within the initial disc. In the third panel of Figure 13 the tail does stretch out to a radius of 4.3, which might be expected since the orbital resonance is with stars at the outer edge of the disc. 

At their longest extents the tails in both Figures 12 and 13 are more like those in Figure 4 (single impulse amplitude of $A = 0.8$) than those of Figure 6 (single impulse amplitude of $A = 0.3$), which emphasizes the cumulative effect of the multiple impacts. However, like low amplitude single impulse cases, the tails of Figure 13 are smooth, with no major clumps. The fourth panel of Figure 12 shows clumps in the form of swallowtail caustics in both tails (e.g., at time 3). This appears to be an example of the mechanism for tidal dwarf formation discussed in the previous subsection. 

Given the number of free parameters in these models (e.g., impact amplitudes, spacing and frequency), or those in more realistic prolonged encounters, it is clear that a very wide variety of tail structures could be produced. Further exploration of these forms is beyond the scope of this paper.

\section{Nonlinear tidal disturbances}

Rather than trying to explore the evolution of ensembles of multiply perturbed stellar orbits, we want to return to some issues remaining from the discussion in Sections 2 and 3, specifically, the unrealistic symmetry of the tidal tails illustrated there. Because p-ellipses have been shown to give good approximations to orbits in the power-law potentials considered here, we do not believe that the fault lies with these orbit functions. It appears that we need a more accurate representation of the collision impulses than provided by equations (5) and (6). These equations are only linear in the factor $q/r_{min}$, and so are not likely to be sufficiently accurate for either strong or prolonged encounters, which will usually be close encounters in both cases. More complex versions of the tidal impulses will give more complex expressions on the right hand side of equation (10). However, this does not pose a great difficulty.

\subsection{Nonlinear Analytic Models}

To consider the effects of nonlinear perturbations we will use the functions of \citet{ge94} for the impulsive velocity disturbance from a Plummer potential in place of equations (5) and (6) above. (\citet{do10} give more accurate representations for various cases. However, the former will suffice for the present purpose of illustrating the effect of nonlinear terms in the ratio of the initial stellar radius to the closest approach distance.) Specifically, we use the velocity impulses given by equations (7a) and (7b) of Gerber and Lamb, with softening length $\gamma = 0$, which can be written as,

\begin{equation}
\Delta v_r = A \eta v_{\phi o} 
\left[ \frac{\cos \left( 2\phi_o \right) - \eta \cos \left(\phi_o \right)}
{1 - 2 \eta \cos \left(\phi_o \right) + \eta^2} \right],
\end{equation}

\begin{equation}
\Delta v_{\phi} = A \eta v_{\phi o} 
\left[ \frac{\sin \left( 2\phi_o \right) - \eta \sin \left(\phi_o \right)}
{1 - 2 \eta \cos \left(\phi_o \right) + \eta^2} \right],
\end{equation}

\noindent where for convenience we adopt the variable, 

\begin{equation}
\eta = \frac{q}{r_{min}}. 
\end{equation}

\noindent With the substitution of these expressions, equation (10) becomes,

\begin{multline}
\frac{\left(\frac{1}{2} + \delta\right) me 
\sin \left( m{\theta}_o \right)}
{\left[ 1 + e \cos \left( m{\theta}_o \right) \right]} = \\
\frac{A' \left( \cos \left( 2{{\phi}_o} \right)
- \eta \cos \left( {\phi}_o \right) \right)}
{ 1 - 2\eta \cos \left( {\phi}_o \right) + \eta^2 -
A' \left( \sin \left( 2{{\phi}_o} \right)
- \eta \sin \left( {\phi}_o \right) \right)},
\end{multline}

\noindent and the orbital solutions are still of the same form as equations (11) Ð (13). That is, equation (11) is unchanged, but equations (12) and (13) for $e$ and $\psi$ become,

\begin{multline}
\left(\frac{1}{2} + \delta\right) me =
\left[\right. A'^2 \left( \cos \left( 2{{\phi}_o} \right)
- \eta \cos \left( {\phi}_o \right) \right)^2 +\\
\left(\frac{1}{2} + \delta\right)^2 m^2 \times\\
\left( 2\eta \cos \left( {\phi}_o \right) - \eta^2 +
A' \left( \sin \left( 2{{\phi}_o} \right)
- \eta \sin \left( {\phi}_o \right) \right) \right)^2
\left.\right]^{\frac{1}{2}},
\end{multline}

\noindent and,

\begin{multline}
\tan \left( 2\phi_o + \psi \right) = 
\left(\frac{1}{2} + \delta\right) m \times\\
\left[ \frac{\left( 2\eta \cos \left( {\phi}_o \right) - \eta^2 +
A' \left( \sin \left( 2{{\phi}_o} \right)
- \eta \sin \left( {\phi}_o \right) \right) \right)}
{A' \left( \cos \left( 2{{\phi}_o} \right)
- \eta \cos \left( {\phi}_o \right) \right) } \right].
\end{multline}

\noindent The expression for the eccentricity in equation (35) is always positive, but can exceed unity.

Equation (35) is a generalization of equation (12). Recall that orbits with eccentricities given by equation (12) were found to be moderately accurate (see Figures 1 and 2). Subsequently, equation (12) was modified with the amplitude dependent coefficient to the cosine term shown in equation (15) to moderate unrealistically large eccentricities in large amplitude orbits. A second modification applied in going from equation (15) to (16) was used to increase the unrealistically low eccentricities of orbits near the backbone (Figure 3). Given the complexity of the perturbation (equations (31) and (32)) it is not surprising that the orbital solutions derived from equation (35) are somewhat less accurate over the whole range of $A'$ and $\delta$ values considered than in the pure tidal case. Adding the extra coefficients of equation (16) to equation (35), is also not as effective in correcting the orbits as it was in the tidal case. However, we can do nearly as well by using slightly modified coefficient. By experimentation we have found the following expression for the eccentricity works well over the relevant parameter range,

\begin{multline}
\left(\frac{1}{2} + \delta\right) me = \\
\left[ A'^2 \left( \left( \frac{1}{1+A'} \right)
 \cos \left( 2{{\phi}_o} \right)
- \eta \cos \left( {\phi}_o \right) \right)^2 \right. +
 \left(\frac{1}{2} + \delta\right)^2 m^2 \times\\
 \left.\left[ 2\eta \cos \left( {\phi}_o \right) - \eta^2 + A' 
 \left( b_3
\sin \left( 2{{\phi}_o} \right)
- \eta \sin \left( {\phi}_o \right) \right) \right]^2
\right]^{\frac{1}{2}}\\
\mbox{with } b_3 = \left(1+A' \right) ^{1+6\delta^2}.
\end{multline}

\noindent Figure 14 shows a comparison between the resulting analytic orbits and numerically integrated orbits. 

\begin{figure}
\centerline{
\includegraphics[scale=0.4]{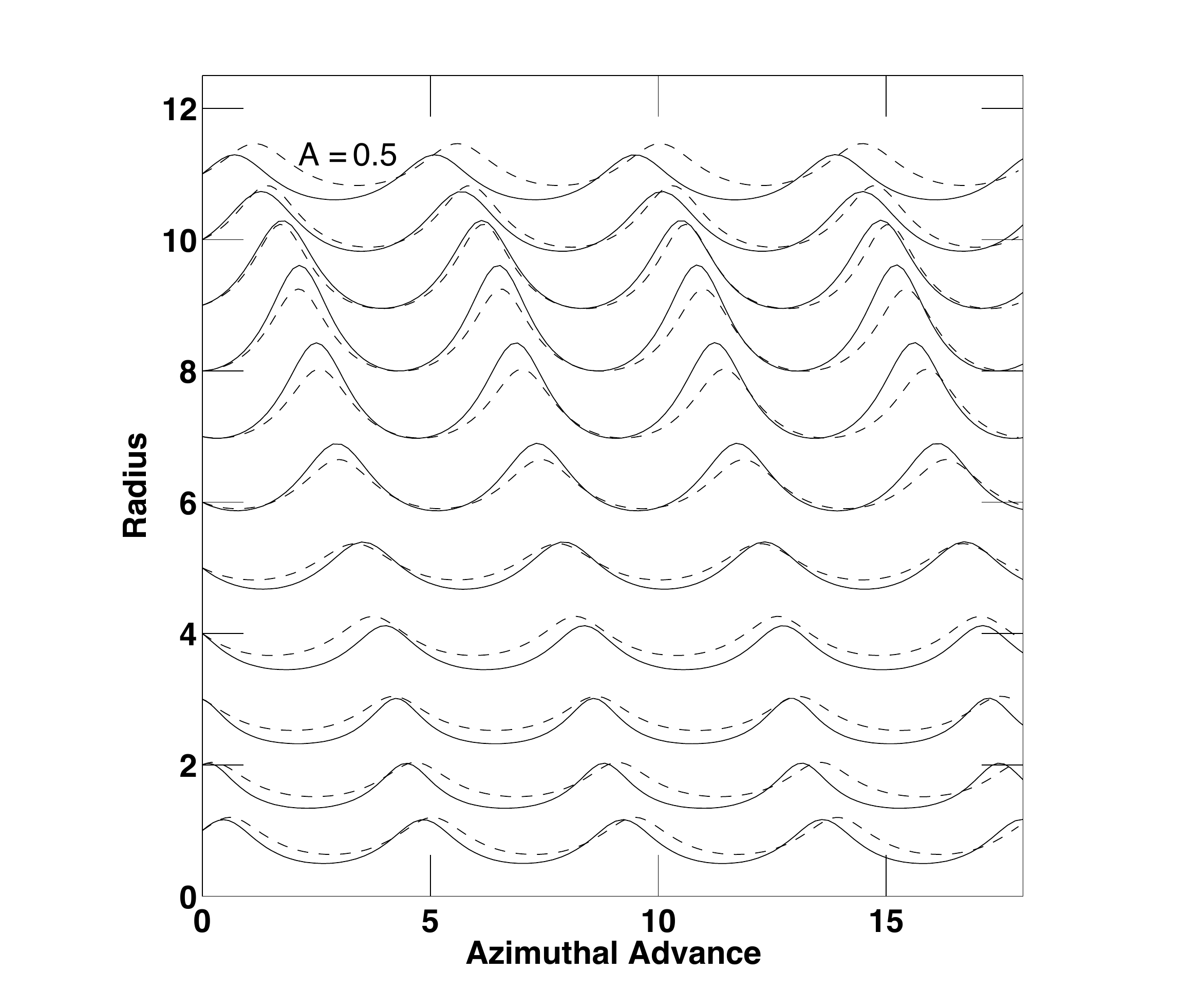}}
\caption{Same as Fig. 1, but with a nonlinear, impulsive disturbance as given by equations (31) Ð (33) with $\eta= 0.10$ when $q = 1.0$.}
\end{figure}

\subsection{Results from the nonlinear model}

The first result from the nonlinear model of the last subsection is a symmetry breaking. In the basic model of Section 2 the amplitude of the tidal perturbation was determined only by the factor $A'$, in the nonlinear model there is an additional dependence on the factor $\eta$ (see equations (31) - (33)). The nonlinear effects are most important for stars with large values of $\eta$, i.e., for small impact parameters and for stars with relatively large initial radii $q$. In the limit of small $\eta$, the nonlinear effects are negligible, and the orbit function is nearly the pure tidal p-ellipse. In this subsection we will look at several example stellar discs, like those discussed in Section 3 for the tidal case. 

Figure 15 shows a series of snapshots for a case similar to that shown in Figure 5, and with $\eta = 0.1$ for stars on the outer edge of the disc. For example, this could represent a case of a companion galaxy flying rapidly past (impulsive collision) a 6 times more massive primary galaxy with a closest approach distance of 10 times the radius of its stellar disc. Despite this relatively large distance of closest approach (in terms of the radius of the small galaxy), the asymmetries generated by the nonlinear perturbation are obvious. The general size and structure of the two tails are not greatly different. However, the caustic waves are more complex in the nonlinear cases than in the tidal cases. By the last times shown the asymmetries have developed into a one-armed form.

Figure 16 shows a lower amplitude case, with $A' = 0.3$ (at $q = 1$). The closest approach distance is also smaller ($\eta = 1/3$), so the nonlinear asymmetries are larger in this case. The bilateral symmetry of the pure tidal part of the disturbance is hardly evident. (Compare to Figure 6.) A series of different caustic waveforms appear (as in some multi-hit cases of Section 4) within the one-armed wave. Interestingly, at some times (e.g., times 3 and 5 in the figure) there are over-dense regions at the end of the arm. Evidently such a seed for the growth of a tidal dwarf galaxy can be quite readily produced in an impulsive, but nonlinear, encounter. The over-density is produced by caustic waves, but via a mechanism that is different than that conjectured in Section 4.2. Note that the mechanism requires a two-dimensional interaction, otherwise the crossing star-streams could be in different planes, and not contribute to a true density enhancement.

\begin{figure}
\centerline{
\includegraphics[scale=0.4]{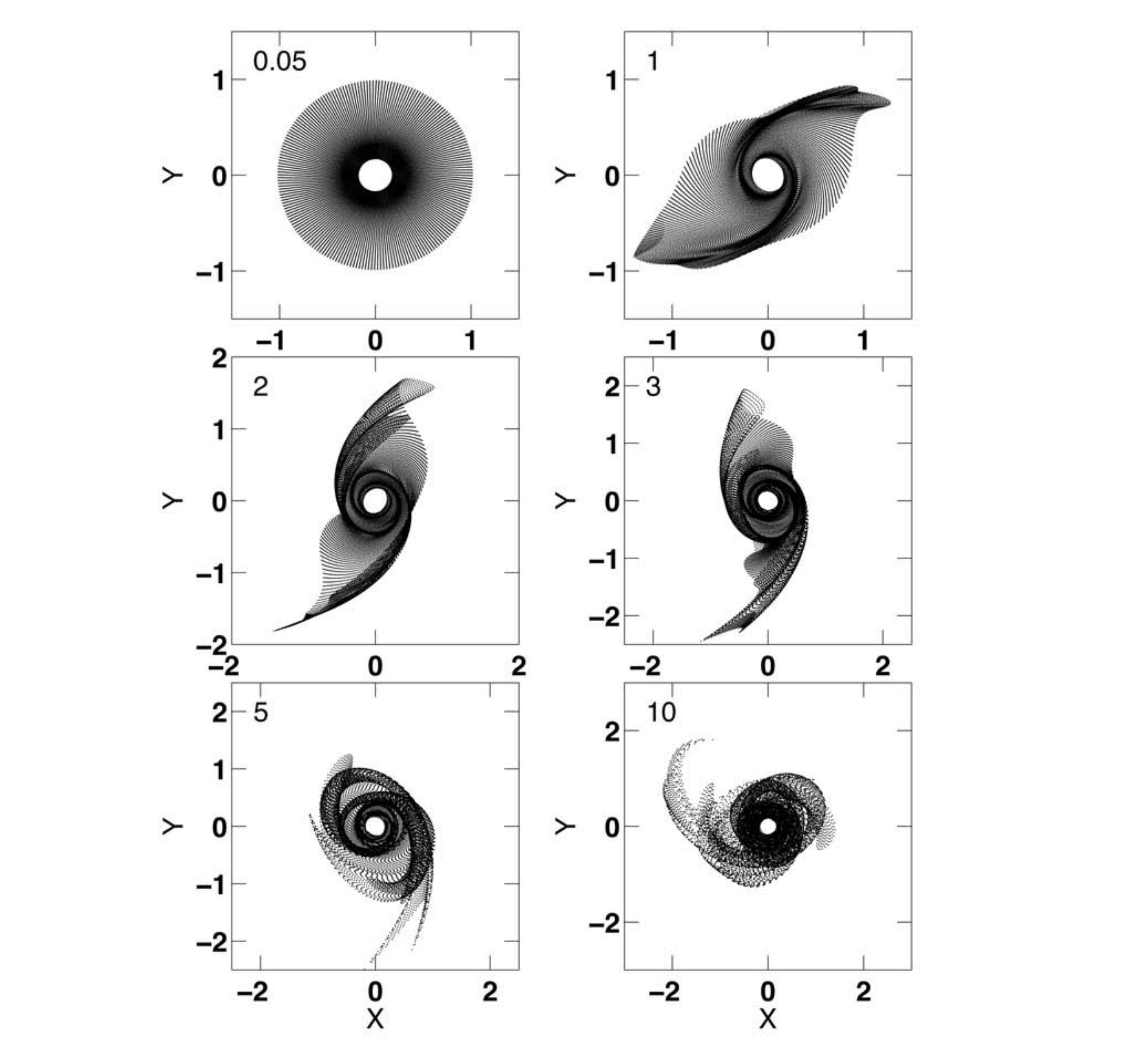}}
\caption{Snapshots of a model in the logarithmic potential with an amplitude of $A' = 0.60$ (at $q = 1.0$), but with a nonlinear, impulsive disturbance as given by equations (31) Ð (33) with $\eta = 0.10$ when $q = 1.0$ (i.e., closest approach at 10 times the initial disc radius).}
\end{figure}

\begin{figure}
\centerline{
\includegraphics[scale=0.4]{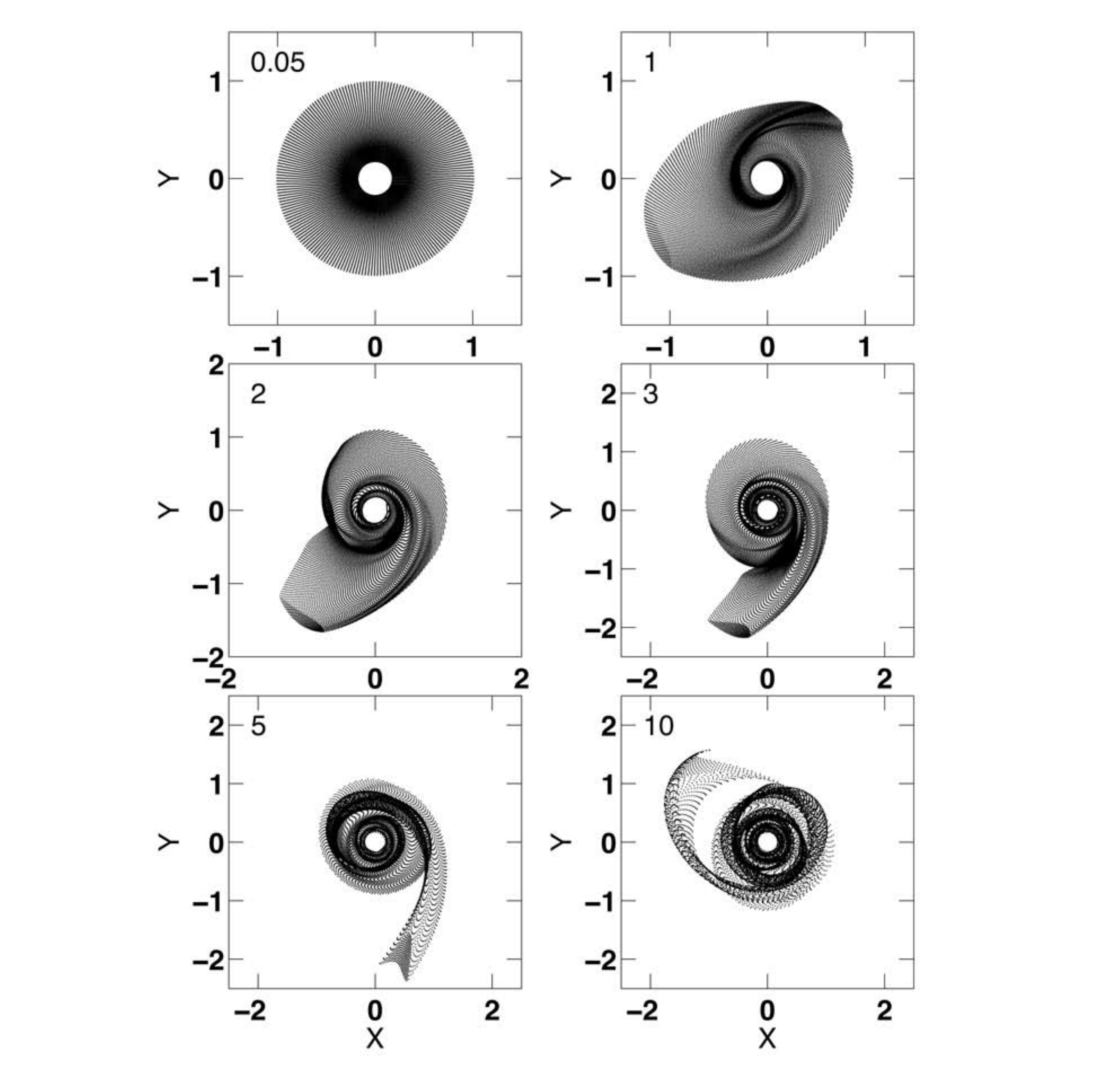}}
\caption{Same as Figure 13, but with $A' = 0.6$ and $\eta = 0.333$ (at $q = 1.0$).}
\end{figure}

High-density zones of converging flows are also common at the base of the tails in Figures 15 and 16. These provide models of a hinge clump. Like many observed hinge clumps (see \citealt{sm10b}), and other features in the nonlinear models these are not generally symmetric. In tidal models they do have a twin on the opposite side of the disc (as seen in Figure 4 for example). Hinge clumps are a recently identified structure, with as yet only a few known examples. The existence of a production mechanism, like that shown in the figure, provides support for the idea that they are a true physical structure, and not simply random clumps of star formation. We will consider more comparisons to observation in the next section.

Figure 17 shows a moderate disturbance ($A' = 0.3$), with a moderate distance of closest approach ($\eta = 1/5$), in the Keplerian potential. Asymmetries like those in Figures 15 and 16 are apparent. The most impressive feature is the long, one-sided tail that develops at late times. 

\begin{figure}
\centerline{
\includegraphics[scale=0.4]{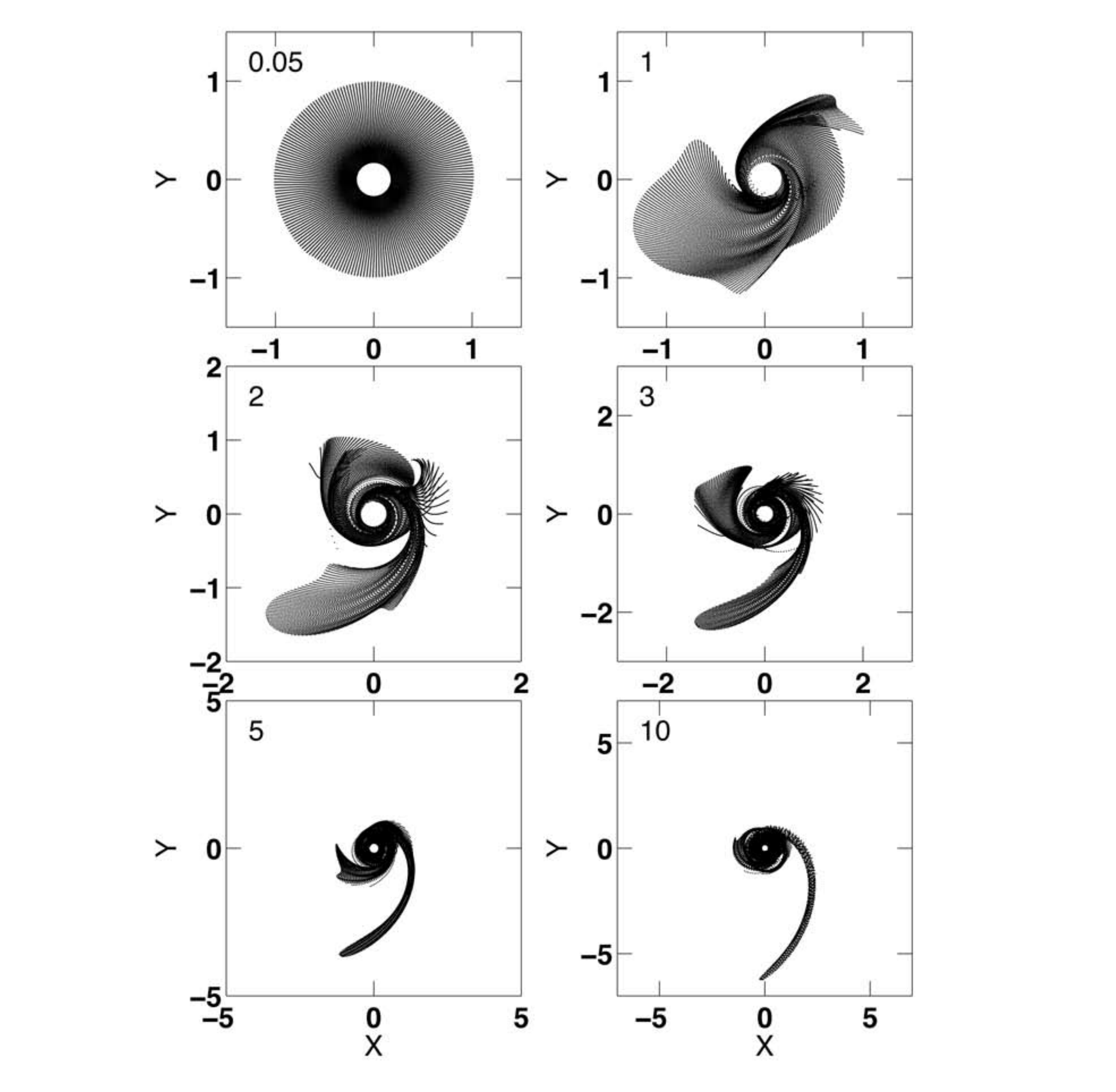}}
\caption{Same as Figure 14, but in the Keplerian ($\delta = 0.5$) potential, and with $\eta = 0.2$ (at $q = 1.0$).}
\end{figure}

\section{Comparisons to observed forms}

In developing the models presented in this paper we have two general goals: 1) to provide simple analytic tools for studying the theory of tidal tails and associated phenomenon in general, and 2) to provide analytic tools to aid in deciphering the collision history of individual observed systems. The previous sections have been occupied with the first goal. In this section we focus on the latter, and specifically, undertake the relatively modest goal of looking for systems that the various models might be applied to, and especially some tentative examples that the single-hit, impulsive models might apply to.

To begin we specify candidate impulsive cases as those with a visible companion that is well separated, and exclude systems that appear to be in prolonged interactions or imminent mergers. Among the former, symmetric cases may be more purely tidal, asymmetric ones nonlinear. For candidates chosen within these restrictions, the following relations predicted by the models can be examined. 1) The tidal dominant cases should have tails that are generally smooth along their length. 2) They may show evidence of a leading edge caustic. 3) In all cases long tails should be associated with large companions, especially in the tidal case where they often do not approach as closely as in the nonlinear case. 4) The longest tails have the least curvature. (This is true even in the Kepler potential.) 5) Irregularities within tails, produced in nonlinear interactions, should be associated with other asymmetries in the disc. 

We note that observed tail characteristics depend on the observational waveband, and whether the observed constituent consists of gas, young stars, or old stars. However, the properties described above are very basic geometric observables that relate quite directly to model characteristics. We also note that the available and adequately resolved observations are not sufficiently extensive to make detailed statistical comparisons (though global statistical properties have been studied recently by \citealt{sm07}, \citealt{sm10a}, \citealt{mo11}). 

With these caveats in mind we will consider example systems that are chosen from the Arp Atlas (\citealt{ar66}), which provides an optically well-resolved sample of sufficient size to contain examples of systems that are somewhat rare even among the rare class of strongly interacting galaxies. The examples discussed below were selected by eye, initially independently by the two authors, with subsequent consultation on systems not chosen by both. This is not a rigorous, objective procedure, but it accomplishes the goal of finding tentative examples and allowing simple tests of the relationships suggested by the models. 

\subsection{Long, straight tails}
One interesting aspect of the impulsive, pure tidal models shown in Figures 4, 5, and to some degree in Figure 7, is the relative straightness of the long tails produced in those cases. (At least in the parts of the tails that lie some distance from the disc.) We expect them to be rare, because a strong perturbation is required to produce a long tail, but an impulsive encounter is needed to produce the simple, symmetric form. In fact, long, symmetric tails are very unlikely, because a strong perturbation generally requires a large companion, and at most a slightly hyperbolic relative velocity. In such cases, dynamical friction can hold the collision partners together, with the formation of a bridge plus counter-tail, rather than two symmetric tails.

These expectations are borne out by the recent
statistical study of the lengths and curvature of tidal tails
by \citet{mo11}.   They found that most tidal
tails have lengths less than three times the diameter of
the disc; only a small fraction have lengths longer than four
disc diameters.   Furthermore, most tails have significant
curvature on the sky, thus intrinsically straight tails
are likely relatively rare.

When observed (doubly or singly), long, straight tails are often assumed to be edge-on. For example, the straightness of the northern tail of Arp 242 (the Mice) may be caused by the disc being edge-on (e.g., \citealt{to72}, \citealt{mi93}, \citealt{bo04}) In general, however, the tendency of tails to stay in their initial plane means this is unlikely unless the parent disc is also observed edge-on. (Not to mention the fact that it is statistically unlikely.) 

\begin{figure*}
\centerline{
\includegraphics[scale=0.5]{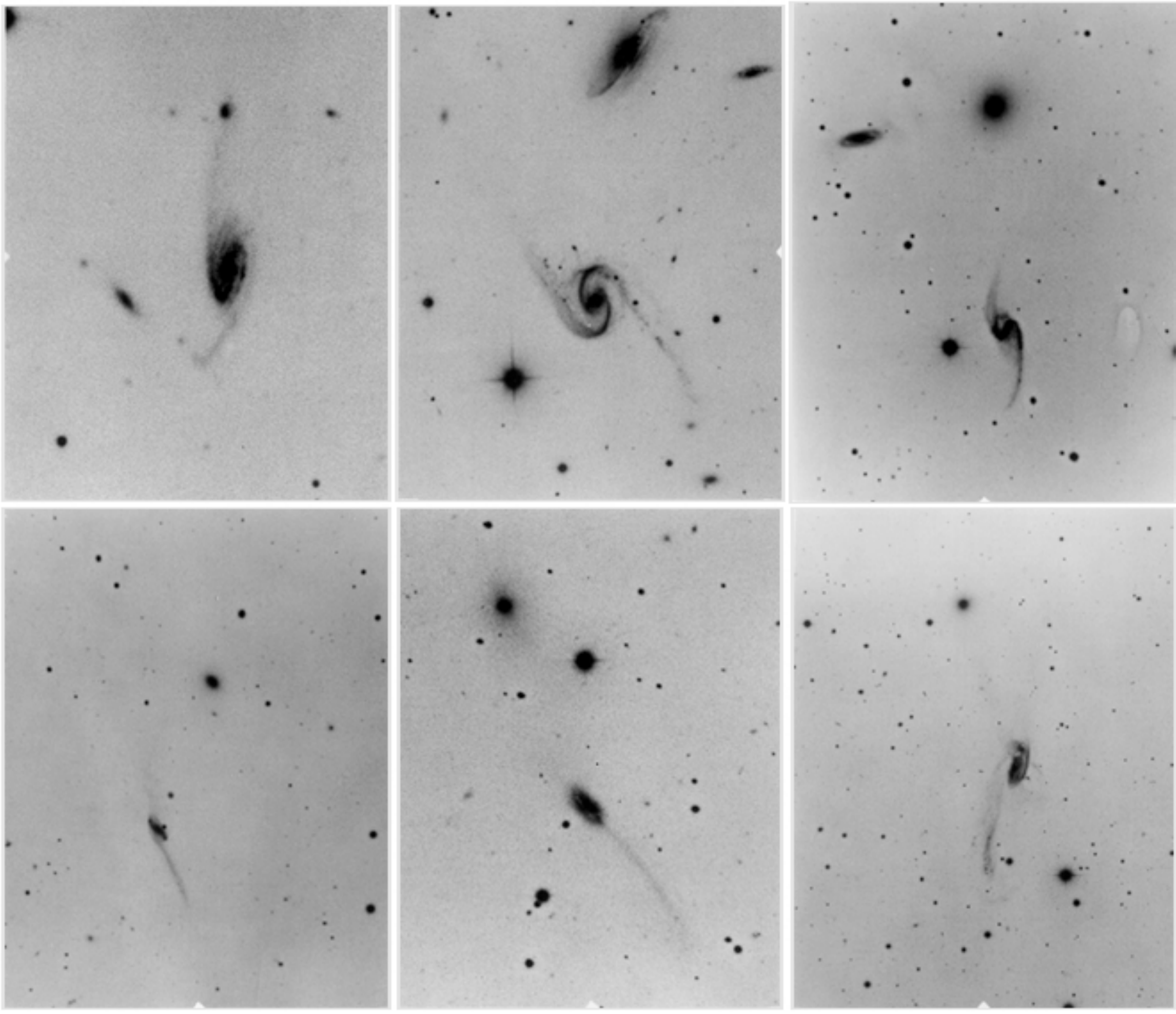}}
\caption{Top row: Examples of Arp galaxies with long straight symmetric tails.
Left to right: Arp 64, 65, and 99.
Bottom row: Examples of Arp galaxies with long, straight, but asymmetric
tails.  Left to right:  Arp 100, 101, and 102. Images from \citet{ar66}}
\end{figure*}

\subsubsection{Symmetric objects}

Nonetheless, the Arp Atlas does have a few long tailed systems that are close to symmetric, including objects: 64, 65, 99, 295, and 296 (see the top row of Figure 18). The disc of Arp 295 does look edge-on, while that of Arp 296 is too small to tell. The discs of the first three objects do not appear edge-on. In all of these cases, except Arp 64, the tailed galaxy has a massive companion, at a distance of at least a couple of disc diameters, consistent with expectations from an impulsive, tidal model. Arp 64 has a ÔhookÕ at the end of one tail, and may have two companions, and so, may be more complex dynamically. In Arp 64, Arp 65 and Arp 99, the tails connect to strong symmetric spirals within the disc. The small fraction of these few long symmetric tails out of the few hundred systems in the Arp Atlas, seems qualitatively consistent with the restricted class of encounters that can produce them. 

\subsubsection{Asymmetric objects}

We can also find apparent examples of long, straight, but single or very asymmetric tails, including Arp objects: 100, 101, 102, and possibly 190, if it is not presently merging (see the bottom row of Figure 18). These tails suggest a very nonlinear disturbance with at least one strong impulse like the models shown in Figure 16 and 17, but on a larger scale, and at somewhat later time. It is true that such singular tails are easier to manufacture in the Kepler potential than in the logarithmic, so in the Arp galaxies they may extend beyond the point where the parent galaxy halo potential begins to fall off. 

It is unfortunate that these particular examples are apparently quite distant, so it is hard to discern caustic structures within their discs. The tails of both Arp 100 and 101 appear smooth, without prominent clumps, yet they are narrow, with blue optical colors and high optical surface brightnesses \citep{sc90}. 

In contrast, the ends of the long straight tails of Arp 102 and 105 are surrounded by a lot of debris, with knots of star formation at the ends (\citealt{sc90}, \citealt{du97}). The base of the Arp 105 tail is narrow and well-defined, while Arp 102 has a broad tail. \citet{du97} suggest that the straightness of the Arp 105 northern tail was caused by the interaction being seen almost edge-on. The Arp 190 tail has star formation near its tip \citep{sc90} and a broader base. Arp 179 does not have an obvious companion galaxy, thus it may be a merger remnant. Many of the long straight tails in the Arp Atlas, like those of the Arp 226, the Atoms for Peace galaxy, are in systems where there is clear evidence that a merger is in progress (e.g., \citealt{to78}). 

A final cautionary example is provided by Arp 188 (the Tadpole), which has a long straight tail. However, Hubble Space Telescope (ACS) imagery seems to affirm the impression from earlier images that the disc of this galaxy has been distorted in three dimensions \citep{tr03}. Extra-planar disturbances and their consequences have not been considered in this paper. In cases when such disturbances are small the model formalism could easily be extended to include epicyclic motions in the vertical direction. 

The fact that we have to look out over large volumes to find examples of this class is suggestive. It must be harder to produce nonlinear disturbances impulsively, than pure tidal disturbances. Nonlinear disturbances will usually be the result of prolonged encounters.

\begin{figure*}
\centerline{
\includegraphics[scale=0.5]{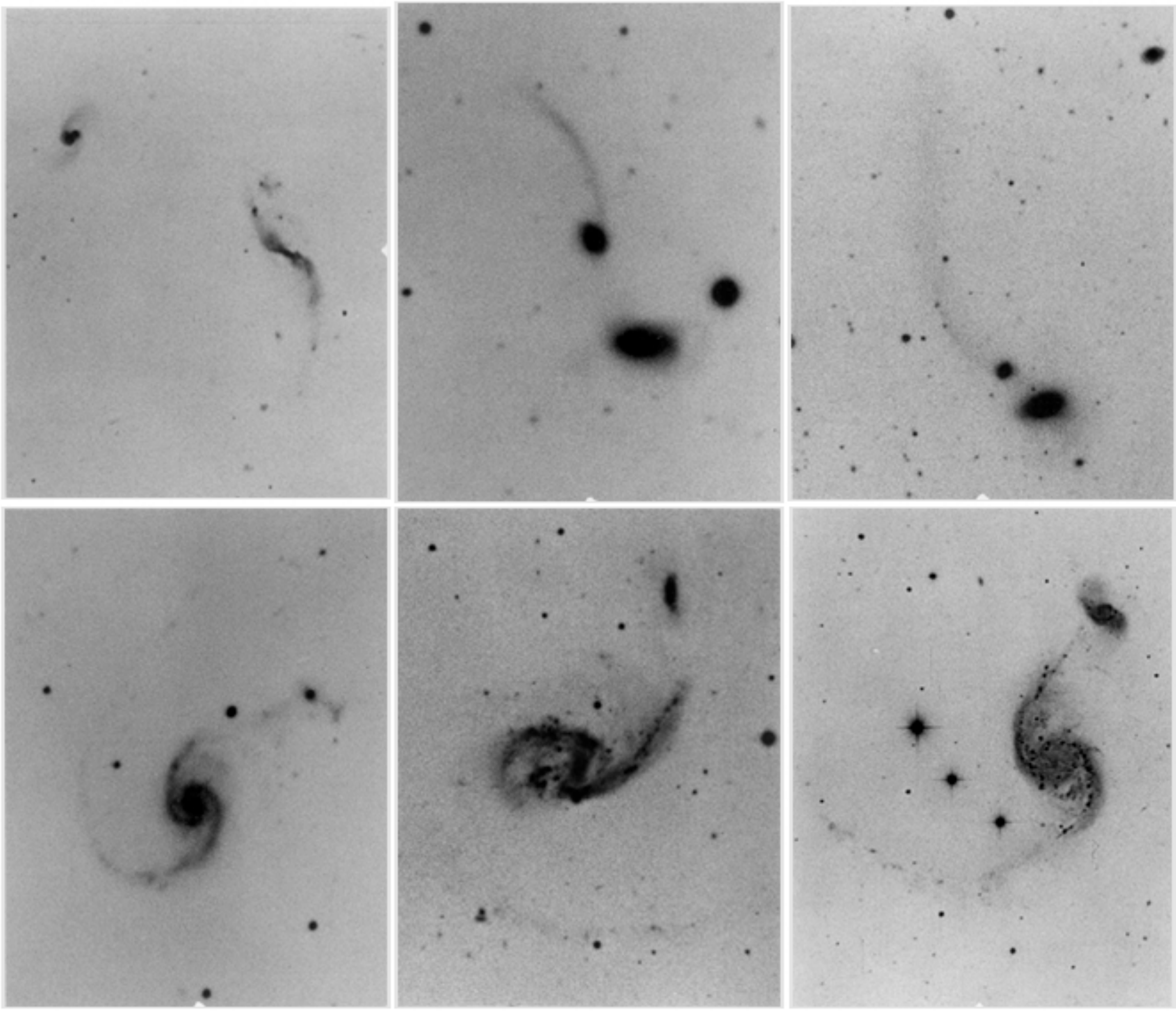}}
\caption{Top row: Examples of Arp galaxies with long but somewhat curved tails.
Left to right:  Arp 33, 173, and 174.
Bottom row: Examples of Arp galaxies with long very curved tails.
Left to right:  Arp 58, 72, and 82. \citet{ar66}}
\end{figure*}

\begin{figure*}
\centerline{
\includegraphics[scale=0.5]{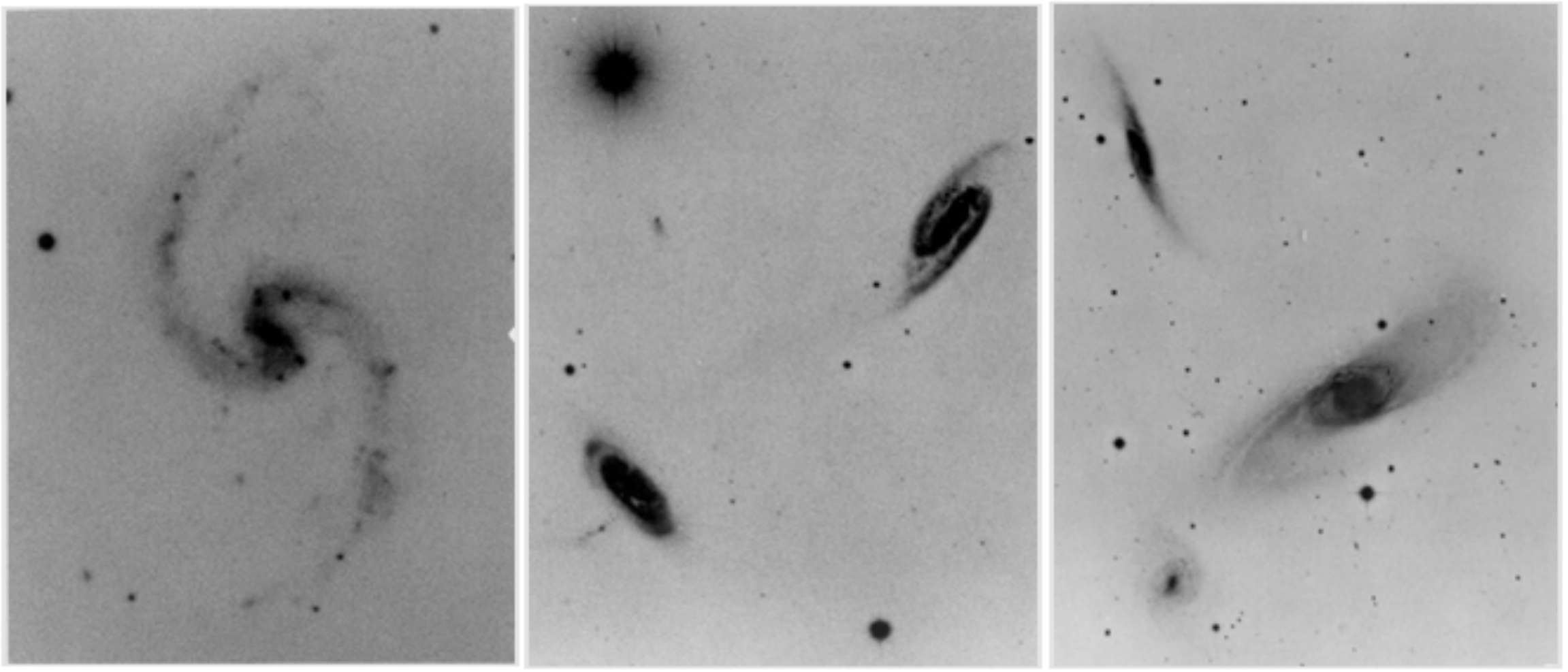}}
\caption{Examples of Arp galaxies with short symmetric tails.  Left to right: Arp 31, 285, 286. \citet{ar66}}
\end{figure*}

\subsection{Long, somewhat curved tails}

If we relax the straightness requirement, and allow somewhat more curved tails, we find more systems, including Arp 33, Arp 173 and Arp 174 (see top row of Figure 19). In the case of the first of these most of the curvature is near the base of the tail. In going from Figure 4 to 6 we see a trend of more curvature at the base of the tail with decreasing perturbation amplitude. The companion in the Arp 33 system appears smaller than the long-tailed object, suggesting a smaller perturbation than in the systems considered above. In the SDSS images the long-tailed galaxy is quite blue, with prominent knots of star formation near the end. The companion also has a nice, symmetric pair of shorter tails, but without bright star formation regions. 

The tailed galaxies in the Arp 173 and 174 systems are both close to larger companions (in projection), so it seems likely that the encounter has been prolonged. In contrast to the Arp 33 tails, the Arp 173 tail is smooth, with relatively red UV/optical colors \citep{sm10a}. The Arp 174 tail is relatively diffuse, also without prominent star forming regions.

Another system with long somewhat curved tails is Arp 96. The tails are faint and diffuse, while the inner spiral arms are quite prominent and curve back on themselves. The tails appear to branch out from the spiral arms at about the location where the arms start curling back inwards. A bright elliptical lies near the end of one of the tails.

\subsection{Long curved tails}

The Arp atlas contains a number of long very curved tails, many of whose parent galaxies seem to be near merging or merged. There are also a number whose companions are more than a couple diameters distant, so merger does not seem imminent. These include Arp objects: 58, 72, 82, 97, 98, 252, and possibly Arp 35, though its disc looks sufficiently disturbed that it may have experienced a recent merger (see bottom row of Figure 19). The results of many numerical merger models at pre-merger times (e.g., \citealt{ba92b}, \citealt{mi98}, \citealt{du04}), and a numerical model for the Arp 82 system in particular (see \citealt{ha07}), teach us that such tails can be formed in prolonged interactions with a bound companion. It takes a considerable time to pull tails out to the observed lengths, yet the companions in these systems are relatively nearby. This is circumstantial evidence for the conclusion that the companions are bound, and have been interacting for some time. These types of tails are not produced in the single impulse models described above. In the multiple hit cases shown in Figures 12 and 13 the interaction is probably not strong enough, or prolonged enough, to produce them.

\subsection{Shorter symmetric or nearly symmetric systems}

Long tails are the most spectacular tidal remnants. Short tails are also of interest, though it is not always entirely clear that they are tidal products. It is possible that some may be spiral waves in relatively low surface brightness discs. On the other hand, strong spiral waves seem unlikely in a low surface density disc, unless there is some external driver. In systems for which there is little data available beyond atlas images ambiguities remain. 

The Arp atlas provides several examples of short, but quite symmetric tails, i.e., in objects: 1, 9, 31, 77, 80, 83, 84, 214, 285, 286, 293 and 316 (see Figure 20). All of these galaxies have possible companions, most with roughly the same redshift, though the apparent mass ratios extend over a large range. Four of these objects (Arp 9, Arp 77, Arp 80 and Arp 285) have obvious bars, so that the `tails' could possibly be bar driven spirals. Five of these objects, and among the most symmetrical, are themselves the smaller companion. These objects are: Arp 83, Arp 84, Arp 286, Arp 293, and Arp 316.

\begin{figure}
\centerline{
\includegraphics[scale=0.75]{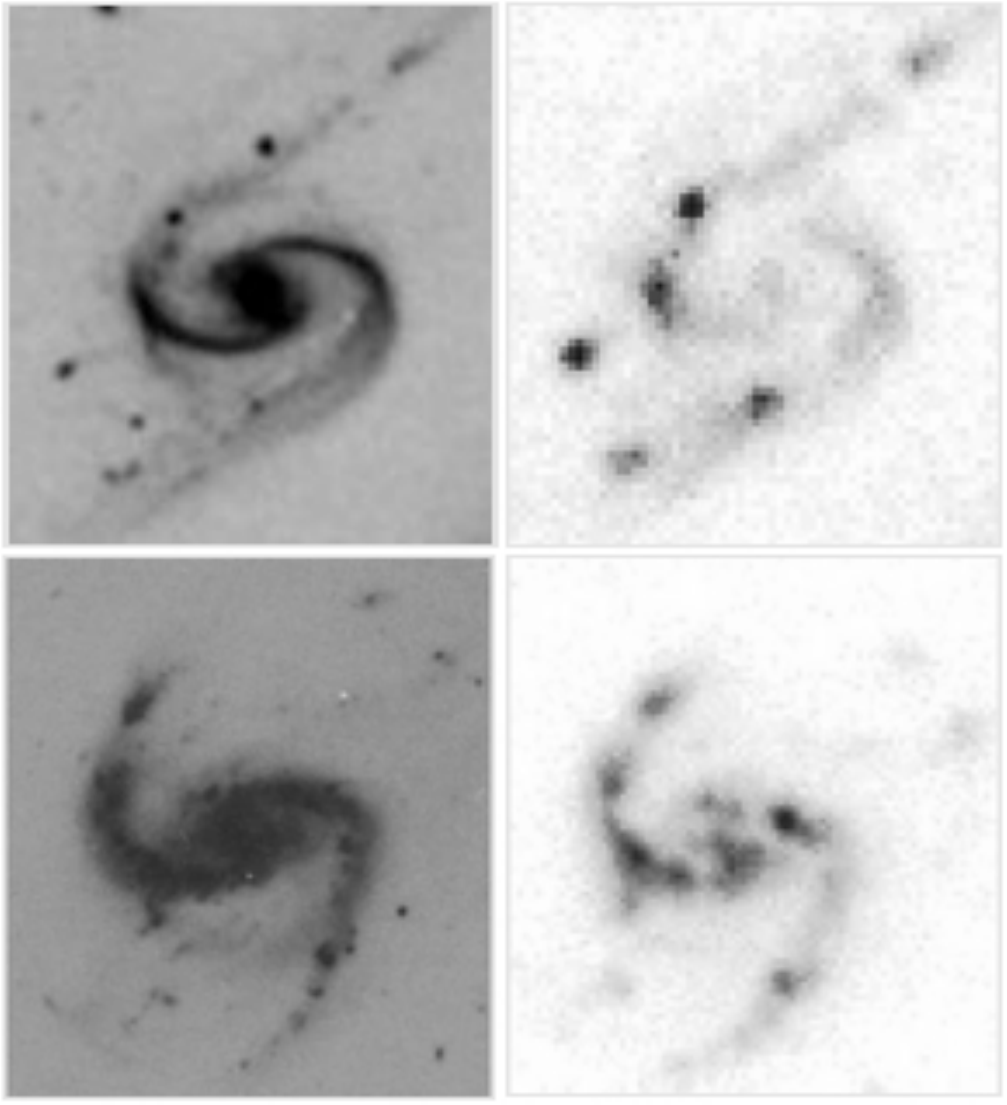}}
\caption{Some example of hinge clumps, as seen in the optical and the UV.
Top row: The western galaxy of Arp 65.   Bottom row: The southern galaxy of Arp 305.
Left column: Optical images from the Arp Atlas.   Right column: GALEX NUV images
from \citet{sm10a}.  North is up and east to the left.
In both systems, two tails extend to the northwest and southeast.
In the Arp Atlas pictures, note the `loop-back' segments from the tails curving back into the inner
regions.  Also note the two UV-bright `hinge clumps' where the `loop-backs'
diverge from the tail.}
\end{figure}

We have previously published numerical models of two of the systems above, Arp 84 (\citealt{ka99}, also see \citealt{pu05}) and Arp 285 (\citealt{sm08}). In both cases the models suggest that the galaxies have experienced a fairly rapid (impulsive) flyby, and are now moving out to apoapse. In the case of Arp 84 the models of Kaufman et al. suggest a closest approach near the outer edge of the primary disc, on a trajectory that is slightly tilted with respect to the plane of the primary disc. The inclination of the companion disc relative to the primary galaxy is similarly modest. The encounter in the simulation was quite impulsive. In the simulation for Arp 285 the relative orbit is in the plane of the primary (tailed) galaxy disc, and the flyby is rapid, so the simulation is very much like the models above. 

Many of these systems bear strong resemblances to the middle panels of Figures 5, and 6, and the timescales of those panels are also about right (e.g., based on the numerical models). Stated differently, the analytic models imply that short, symmetric tails appear shortly after impulsive, but dominantly tidal encounters, and these observational and numerical examples are in accord with those correlations.

\subsection{Swallowtails and hinge clumps}

As discussed in Section 3 and as illustrated in Figure 10,
the models sometimes produce swallowtail caustics
near the base of tidal tails. Intersecting streams of gas
may trigger star formation at these locations, perhaps accounting
for the `hinge clumps' observed in some interacting systems.
Two galaxies that match the basic morphology of Figure 10
are shown in Figure 21: the western galaxy in Arp 65 (top row)
and the southern galaxy in Arp 305 (bottom).
The left panels show expanded
views of the Arp Atlas images of these two galaxies, while
the panels on the right display matching GALEX NUV images.
In the Arp Atlas image, as discussed
by \citet{ha09}, the Arp 305 galaxy has an ocular structure,
with two fairly straight tails coming out of the ends of the `eye'.
Partway along the tails, luminous knots of star formation are seen
(`hinge clumps') on the GALEX image.
Near the hinge clump in the southern tail,
an optical loop is seen in the Arp Atlas picture
that arcs back into the main galaxy.
A similar, but less prominent, loop-back is also seen
near the northern hinge clump.
This morphology resembles the model in Figure 10.
 
A similar structure is seen in the western galaxy
of Arp 65, except that the inner ocular is not
completely closed, being an open symmetric spiral.
The two hinge clumps, seen
to the northeast and southwest of the nucleus,
lie near the location where `loop-backs'
separate from the tail in the Arp Atlas picture.
These hinge clumps are also seen in
the Spitzer 8 micron images of \citet{sm07}.
 
Another system that may have a similar structure
is Arp 99, which also has long straight tails like
Arp 65.  A published H$\alpha$ map shows numerous
knots of star formation along the tails, with
particularly luminous clumps at the bases of the
two tails (\citealt{ig01}).
The Arp Atlas image of this galaxy also suggests
`loop-backs' near these hinge clumps.
Hinge clumps are also seen in galaxies with
long curved tails.  For example,
Arp 82 also has a luminous hinge clump at the
base of its northern tail (\citealt{ha09}).  Arp 58 may have a similar
structure, judging from the Arp Atlas photograph and
an archival GALEX UV image.   Although more disturbed,
UV and optical images of Arp 72 (\citealt{sm10a},
\citealt{ar66}), and infrared and optical images of Arp 240 \citep{sm07}, suggest the same basic structure.

\subsection{Observational comparison summary}

About 5-10\% of the objects in the Arp atlas may provide examples of interacting systems that can be described by the impulsive analytic models above. This percentage might be considerably increased when the models for multiple or prolonged flyby encounters are included. This is about what we would expect given that the models of this paper are restricted to planar, flyby collisions, and exclude mergers (in progress or completed), and three-dimensional encounters. 

The sample systems also seem to be in general agreement with correlations suggested by the analytic models, noted at the beginning of this section, and discussed in the preceding subsections. We conclude by mentioning a couple more. Firstly, we emphasize that most of the long tails of Section 6.1 and the short tails of Section 6.4 are quite smooth along their length, with the exception of some knots of star formation. The barred objects discussed in the previous subsection are somewhat less smooth than the unbarred systems. This makes sense in the context of interference effects between tidal and bar-driven perturbations. The shorter tails tend to have higher surface brightnesses, and some appear to have quite sharp leading edges, as we would expect for azimuthal caustics (e.g., Arp 83, Arp 285, Arp 293, and perhaps Arp 316). We would reemphasize that among the symmetric tails, the shorter ones have more curvature, and the long ones tend to be straighter. In sum, the models provide fair representations of the basic phenomenology, and the observations of those systems do not obviously contradict model properties. However, the comparisons here are only qualitative, and given the limited data it is hard to be sure that all the examples satisfy the approximations on which the models are based. 

\section{Summary and Conclusions}

The p-ellipse functions describe orbits of stars in power law potentials very naturally. Because they allow large radial excursions, tidal tails and related structures are also naturally produced with these orbit functions. These model tidal structures appear to match the observations well in systems where the model approximations (e.g., of an impulsive planar flyby) may obtain. We conclude with a summary of our main results.

1. The p-ellipses provide simple, yet accurate approximations to impulsively perturbed orbits. However, at moderate to large eccentricities the best fit is achieved with a change in the p-ellipse eccentricity from that in the simplest solution of equation (12),  to the modified version of equation (16).

2. One aspect of this fit, the result from equation (12) that $e \approx A'$, is an especially simple result, if not directly observable.

3. The form of the tidal perturbation suggests that tails form around a backbone at a specific azimuth. The p-ellipse orbits for stars on the backbone provide a very simple model for the evolution of the simplest tails. This analysis yields a formula for how the maximum length of the tail depends on the amplitude of the tidal perturbation. This result could be estimated from angular momentum conservation without reference to the p-ellipse. However, the dependence on the halo potential is a simple result of using the p-ellipse equation. Moreover, as shown in Section 3 the p-ellipse orbits can also reveal the structure of the whole tail as a function of initial position and time. (Also see \citealt{ly10b}.)

4. The generic evolution of the orbit ensembles also supports some observable generalizations (and accords with some simulation results). For example, tails produced by tidal impulses are smooth, symmetric, but generally develop an azimuthal caustic region of enhanced density on their leading edge, after traveling $90^{\circ}$ from their initial azimuth. Long tails are very straight. Nonlinear (not purely tidal) impulses can produce higher order caustics, and less regular tails, as can multiple tidal impulses. In both cases, detailed tail structure depends on the halo potential (assumed rigid in these analytic models). 

5. The p-ellipse approximation provides the basis for interpretations of a variety of other tidal phenomena, including ocular rings (radial caustics corresponding to the angular caustics in the tails), and hinge clumps (as higher order caustics located near the base of tails). The orbit formulae can also be used to derive evolutions and scalings for these structures via analysis of the caustics. 

6. In cases of two or more impulsive encounters, the single impulse results can be extended to produce estimates of tail length in prograde versus retrograde cases, as a function of the form of the potential. Two impulsive encounters (e.g., viewed as a simple approximation to an encounter of finite duration) provide the simplest model for the prograde/retrograde dichotomy. 

7. In the case of two encounters, the map of the second perturbation across the disc interacts with the evolved velocities from the first perturbation in a way that can be viewed as an interference pattern. On the (second) backbone an especially strong perturbation can result from this interference at a specific location. This could provide the seed for the formation of a tidal dwarf galaxy, as discussed in Section 4.2. Further work is needed on this conjecture. Nonlinear or prolonged interactions can also generate high order caustic density enhancements that could seed tidal dwarf formation.

8. The most general case considered in this paper is that of multiple nonlinear impulses (see appendix). We have not investigated this case, but the discussion in the appendix provides a template for studying it with any nonlinear perturbation function.  

More generally, the p-ellipse orbits provide a useful tool for furthering our understanding of a wide range of phenomena in colliding or otherwise disturbed galaxies. Previous studies have considered the properties of the p-ellipses, and their unusual accuracy in approximating galaxy orbits (S06, \citealt{ly10a}). The applications above help demonstrate how the family of functions may prove useful in many problems in galaxy dynamics.

\section*{Acknowledgments}

We are grateful to D. Lynden-Bell and an anonymous referee for helpful comments. This research has made use of the NASA/IPAC Extragalactic Database (NED), which is operated by the Jet Propulsion Laboratory, Caltech, under contract with NASA.  It has also made use of NASA's Astrophysics Data System.

\bibliographystyle{mn2e}

\section{Appendix: Equations for the case of two impulses}

\subsection{Orbital equations in the pure tidal case}

Like the first impulse, to derive the orbit after a second impulse we add the component velocity changes to the current velocity, assume the position is unchanged during the impulse, and use these quantities to determine a new (rotating) p-ellipse orbit. In Section 2 we were able to use the fact that the form of the p-ellipse velocity ratio is generally the same as that of a circular velocity plus the tidal impulse, to derive a very simple orbital solution. With little more effort this can also be done after a second impulse. The velocity equation ratio analogous to equation (10) can be written,

\begin{multline}
\frac{\left(\frac{1}{2} + \delta\right) m e_{II} 
\sin \left( m{\theta}_{II} \right)}
{\left[ 1 + e_{II} \cos \left( m{\theta}_{II} \right) \right]} = 
\frac{v_r}{v_\theta} = \frac{v_{r3}}{v_{\phi 3}} =\\
\frac{v_{r2} + A'_{II} v_{\phi 2} \cos \left( 2\left(\phi_2 - \phi_c \right) \right)}
{v_{\phi2} \left[ 1 - A'_{II} \sin \left( 2\left(\phi_2 - \phi_c \right) \right) \right]}.
\end{multline}

\noindent The terms on either side of the first equality represent the p-ellipse orbit immediately after the second impulse. It has an eccentricity $e_{II}$, and its angular coordinate is $\theta_{II}(t)$. The final equality gives the immediate post-impulse radial velocity in the numerator, and the azimuthal velocity in the denominator. 

\textit{Henceforth, we use the subscripts 0, 1, 2, and 3 to denote the radius and azimuthal positions and the corresponding velocity components at the following times: just before the first impulse, immediately after the first impulse, just before the second impulse, and immediately after the second impulse, respectively. We use subscript I for parameters of the first p-ellipse orbit (between the two impulses), and subscript II for p-ellipse parameters after the second impulse.} The angle $\phi_c$ gives the direction of the second impulse, relative to the direction of the first impulse, which is at zero azimuth. The amplitude $A'_{II}$ will generally have a form like that given by equations (7) and (9), but with a different value.

Because the orbits are not circular before the second impulse, the numerator of the last term in equation (38) has a term, $v_{r2}$, which is absent in equation (10). However, obtaining a solution like that of equations (11) - (13), is not much more complicated. We can begin with a very similar form for the initial value of the p-ellipse azimuth after the second impulse, $\theta_{IIo}$,

\begin{equation}
m\theta_{IIo} = 2\left(\phi_{2o} - \phi_c \right) 
+ \frac{\pi}{2} + \psi.
\end{equation}

\noindent Then we can use the conditions that the numerators and denominators (radial and azimuthal velocities) in equation (38) are equal to get,

\begin{multline}
e_{II} = \frac{A'_{II}}{\left(\frac{1}{2} + \delta\right) m} \times\\
\left[ \left(\frac{1}{2} + \delta\right)^2  m^2
\sin^2 \left( 2\left(\phi_2 - \phi_c \right) \right) + \right.\\
\left.\left( \cos \left( 2\left(\phi_2 - \phi_c \right) \right)
+ \frac{v_{r2}}{A'_{II} v_{\phi2}}
\right)^2 \right]^\frac{1}{2},
\end{multline}

\noindent and,

\begin{multline}
\tan \left( 2\left(\phi_2 - \phi_c \right) + \psi \right) = \\
\frac{\left(\frac{1}{2} + \delta\right) m \sin \left( 2\left(\phi_2 - \phi_c \right) \right)}
{\left[ \cos \left( 2\left(\phi_2 - \phi_c \right) \right) + \frac{v_{r2}}{A'_{II} v_{\phi2}} \right]},
\end{multline}

\noindent like equations (12) and (13). Note that the extra terms from equation (16) could also be included in equation (40) with the expectation of similar gains in accuracy.

Like all overlapping wave systems, we expect the orbital phases produced by the two impulses, to generate an interference pattern. This complicates the structure of caustics, for example, or at least generates perturbations to the stronger ones. Nonetheless, we will see in the next few subsections that we can derive some interesting results from the solution above.

\subsection{Amplitude of the second impulse}

We continue by looking at the amplitude $A'_{II}$ in more detail. This amplitude is an initial condition of the second p-ellipse. Following the example of equations (7) and (9) we can write,

\begin{equation}
A'_{II} = A_{II} \left( \frac{q_2}{r_{min2}} \right) =
\frac{2GM_c}{r_{min2} V_2 v_{\phi o}}
\left( \frac{q_2}{r_{min2}} \right).
\end{equation}

\noindent Again we use the subscript Ô2Õ to denote values at time $t_2$ just before the second impulse. For example, $V_2$ is then the relative velocity, $r_{min2}$ is the radius of the second close approach, and $q_2 (= r_2)$ is the radius of the representative star at that moment. 

Define the factor 

\begin{equation}
k_A = \frac{A_{II}}{A_I},
\end{equation}

\noindent which compares the result of changes in the collision parameters $r_{min}$, $V$, or even $M_c$ (in the case of two companions) between the two impulses. If the second impulse is simply an approximation to an interaction of finite duration, then we expect that $k_A \approx 1$. If it is due to a second encounter, then since the relative velocity and distance of closest approach probably decrease due to dynamical friction, $k_A > 1$.  For encounters with two different companions, values of greater or less than one are possible. With the $k_A$ factor equation (43) can be used to eliminate $A_{II}$ in favor of $A_I$ in Equation (42).

None of the factors in equation (42) depend on azimuth except $q_2$, the initial radius at the time of the second impulse. This dependence is given by the p-ellipse equation (equation (1) or (2)). I.e., 

\begin{multline}
\frac{q_2}{r_{min2}} = \frac{q_2}{q_1} \frac{q_1}{r_{min2}} =
\frac{q_1}{r_{min2}}
\left[ \frac{1+e_I \cos\left(m\theta_{I1}\right)}
{1+e_I \cos\left(m\theta_{I2}\right)} \right]
^{\left(\frac{1}{2} + \delta\right)}\\
=  \frac{q_1}{r_{min2}}
\left[ \frac{1-A'_I \sin\left(2\phi_o\right)}
{1+A'_I \cos\left(m\theta_{I2}\right)} \right]
^{\left(\frac{1}{2} + \delta\right)},
\end{multline}

\noindent Equation (10) was used to obtain the last equality. We note that $\theta_{I2}$, for example, is defined as the azimuth of the first p-ellipse at the time just before the second impulse. 

Substituting equations (44) into equation (42) and simplifying yields,

\begin{multline}
A'_{II} = k_A  A_I  \left[ \frac{1-A'_I \sin\left(2\phi_o\right)}
{1+A'_I \cos\left(m\theta_{I2}\right)} \right]
^{\left(\frac{1}{2} + \delta\right)}
\left(  \frac{q_1}{r_{min1}} \right)
\left(  \frac{r_{min1}}{r_{min2}} \right)\\
= k'_A  A'_I  \left[ \frac{1-A'_I \sin\left(2\phi_o\right)}
{1+A'_I \cos\left(m\theta_{I2}\right)} \right]
^{\left(\frac{1}{2} + \delta\right)},
\end{multline}

\noindent where the last expression uses equation (9) and the definition,

\begin{equation}
k'_A = k_A \frac{r_{min1}}{r_{min2}},
\end{equation}

The maximum (minimum) value of $A'_{II}$ is obtained when $\sin (2 \phi_o) = -1 (1), \cos(m \theta_{I2}) = -1 (1)$, and $e_I \approx A'_{II}$ (equation (12)). Those extremal values are,

\begin{eqnarray}
\nonumber
A'_{IImax} = k'_A A'_I  
\left( \frac{1+A'_I}{1-A'_I} \right)^{\left(\frac{1}{2} + \delta\right)},\\
\mbox{and,  }
A'_{IImin} = k'_A A'_I  
\left( \frac{1-A'_I}{1+A'_I} \right)^{\left(\frac{1}{2} + \delta\right)},
\end{eqnarray}

In the case where $k'_A \approx 1$, and $A'_I$ is small (e.g., $< 0.1$), then $A'_{IImax} \approx A'_I$. 

$A'_{II}$ is the net amplitude after the two impulses, so this result and the corresponding result for $n$ low amplitude encounters, $A'_{n,max} \approx A'_1 (1 +2nA'_1)$ confirm the conclusion that it would take a very prolonged encounter with a low mass companion to significantly perturb the primary.

When $A'_{I}$ exceeds values of about 0.2 then the values of $A'_{IImax}$ begin to grow more nonlinearly. For values of $A'_{I}$ approaching 0.5, $A'_{IImax}$ approaches or exceeds unity depending on the value of $\delta$. If the approximation of equation (16) was used to obtain $A'_{I}$, then the amplification would be even greater. Thus, the second impulse can drive a very nonlinear increase (or decrease) in the perturbation amplitude, for orbits with the appropriate phases. 

\subsection{Maximal excursions}

Equally interesting is the largest radial excursion of the stellar orbit following a second impulse. We begin by noting that equations (17) and (18) apply to the second p-ellipse (generated by the second impulse), as well as the first. In particular, in the notation of this section, the apoapse radius is given by,

\begin{equation}
\frac{r_{ap}}{q_2} = 
\left(\frac{1+A'_{II}}{1-A'_{II}} \right)
^{\left(\frac{1}{2} + \delta\right)},
\end{equation}

\noindent with $q_2$ and $A'_{II}$ given by equations (44) and (45), and assuming $e_{II} \approx A'_{II}$. (The more accurate approximation of equation (18) would clutter the already complex equations below, without qualitatively changing the conclusions.)

Under the resonant conditions that led to equation (47), and when $k'_A = 1$, equation (45) yields,

\begin{equation}
A'_{II} = A'_I
\left(\frac{1+A'_{I}}{1-A'_{I}} \right)
^{\left(\frac{1}{2} + \delta\right)},
\end{equation}

\noindent and

\begin{multline}
\frac{r_{ap}}{q_1} = 
\left[ \frac{\left(1-A'_{I}\right)^{\left(\frac{1}{2} + \delta\right)} +
A'_I \left(1-A'_{I}\right)^{\left(\frac{1}{2} + \delta\right)}}
{\left(1-A'_{I}\right)^{\left(\frac{1}{2} + \delta\right)} -
A'_I \left(1-A'_{I}\right)^{\left(\frac{1}{2} + \delta\right)}} \right]
^{\left(\frac{1}{2} + \delta\right)}\\
\times \left(\frac{1+A'_{I}}{1-A'_{I}} \right)
^{\left(\frac{1}{2} + \delta\right)}.
\end{multline}

\noindent To first order in the amplitude, the apoapse radius is, 

\begin{equation}
\frac{r_{ap}}{q_1} = 
1 + 4\left( \frac{1}{2} + \delta \right) A'_I,
\end{equation}

\noindent which is about twice the small amplitude increase given by equation (18) (at the same level of approximation).

This shows that in optimal circumstances, the apoapse radius of a stellar orbit can grow geometrically with successive impulses, or prolonged disturbances. The dependences on $\delta$ show the dependence on the form of the potential. I.e., the effect is roughly doubled in Keplerian potentials relative to flat-rotation-curve potentials. 

\subsection{Nonlinear tidal disturbances}

In this subsection we extend the mathematical results of Section 5.1 for nonlinear impulses to the case of two impacts. As in the tidal case, consideration of multiple impacts introduces additional parameters, e.g., the amplitude and impact parameter of the second interaction. We will not extend the sample disc evolution investigations of Section 5.2, since exploration of this larger parameter space is beyond the scope of this paper. The goal here is primarily to record the equations for future use, either with the disturbance functions of equations (31) and (32), or with other forms.  

We begin with the velocity matching equations, analogous to equation (48), with the equations for the radial and azimuthal velocities, but written separately rather than in a ratio, 

\begin{multline}
\left(\frac{1}{2} + \delta\right) m e_{II} 
\sin \left( m{\theta}_{3} \right) =\\
\frac{v_{r2}}{v_{\phi 2}} +
 A'_{II} \left( \cos \left( 2\left(\phi_2 - \phi_c \right) \right) -
 \eta_{II} \cos \left(\phi_2 - \phi_c \right)  \right),
\end{multline}

\begin{multline}
e_{II} \cos \left( m{\theta}_{3} \right)  = 
-2 \eta_{II} \cos \left(\phi_2 - \phi_c \right) + \eta^2_{II}\\
- A'_{II} \left( \sin \left( 2\left(\phi_2 - \phi_c \right) \right) - 
\eta_{II} \sin \left(\phi_2 - \phi_c \right)  \right).
\end{multline}

\noindent where the subscript notation follows the conventions of Section 4. That is, subscript 2 denotes the values of quantities at a time just before the second impulse, subscript 3 denotes quantity values immediately after the second impulse, and subscript $c$ is used for the azimuth of the companion galaxy at the time of the second impulse (or closest approach). The subscript $II$ denotes parameters of the new p-ellipse orbit after the second impulse, e.g., the eccentricity. The quantity $\eta_{II}$ equals the radius of the star at the time of the second impulse divided by the minimum approach radius of the companion.  As in Section 5 this is the nonlinearity parameter. 

We assume that we know the parameters of the first p-ellipse orbit, generated by the first impulse, and the time between impulses (and also the azimuth $\phi_c$).  Then we know the subscript 2 quantities, i.e., the positions and velocities to which the stars have advanced along their p-ellipse orbits during the time between impulses. In particular, the velocity ratio in equations (52) and (53) can be written in terms of the starÕs azimuth before the second impulse and the eccentricity of its (first) p-ellipse orbit (as in equation (4)),

\begin{equation}
\frac{v_{r2}}{v_{\phi 2}} = 
\frac{\left(\frac{1}{2} + \delta\right) m e_{I} 
\sin \left( m{\theta}_{2} \right)}
{1 + e_{I} \cos \left( m{\theta}_{2} \right)}.
 \end{equation}

\noindent As in Section 4 we develop the solution for the second p-ellipse orbit by adopting the following expression for its initial azimuth (see equation (39)), 

\begin{equation}
m \theta_3 = 2 \left( \phi_2 - \phi_c \right) + \frac{\pi}{2} + \psi.
 \end{equation}

\noindent Then, the eccentricity of the new p-ellipse orbit can be derived by squaring and adding the two velocity equations (52) and (53) to eliminate the variable $\theta_3$. The result is, 

\begin{multline}
\left(\frac{1}{2} + \delta\right)^2 m^2 e^2_{II} = \\
\left[ \frac{v_{r2}}{v_{\phi 2}} +
 A'_{II} \left( \cos \left( 2\left(\phi_2 - \phi_c \right) \right) -
 \eta_{II} \cos \left(\phi_2 - \phi_c \right)  \right) \right]^2\\
 + \left[ \right. -2 \eta_{II} \cos \left(\phi_2 - \phi_c \right) + \eta^2_{II}\\
- A'_{II} \left( \sin \left( 2\left(\phi_2 - \phi_c \right) \right) - 
\eta_{II} \sin \left(\phi_2 - \phi_c \right)  \right)
\left.  \right]^2.
\end{multline}

\noindent An equation for the phase $\psi$ can then be derived from equation (52) after substituting equation (55),

\begin{multline}
\cos \left( 2\left(\phi_2 - \phi_c \right) + \psi \right) = \\
\frac{\frac{v_{r2}}{v_{\phi 2}} +
 A'_{II} \left( \cos \left( 2\left(\phi_2 - \phi_c \right) \right) -
 \eta_{II} \cos \left(\phi_2 - \phi_c \right)  \right)}
 {\left(\frac{1}{2} + \delta\right) m e_{II}},
\end{multline}

\noindent and where equation (54) can be used for the velocity ratio. These equations complete the second impulse solution.

Two impulse models may not accurately represent interactions of significant duration, but they do illustrate some of the qualitative forms that can be generated in such encounters, as discussed in Section 4. The equations of this appendix can be readily generalized to describe a sequence of nonlinear encounters, to create more realistic semi-analytic models of prolonged encounters. Alternate forms can also be used in the right hand sides of equations (52) and (53) to represent disturbances different than the impulsive Plummer potential types used here.

\bsp
\label{lastpage}
\end{document}